\documentclass[a4paper, 11pt]{article}
\usepackage{jheppub}

\preprint{MCnet-15-11, CP3-15-16}

\title{\boldmath Interference effects for $H\to WW\to \ell\nu q\bar{q}'$ and $H\to ZZ\to \ell\bar{\ell}q\bar{q}$ searches in gluon fusion at the LHC}

\author[a]{Nikolas Kauer,}
\author[a]{Claire O'Brien}
\author[b]{and Eleni Vryonidou}
\affiliation[a]{Department of Physics, Royal Holloway, University of London, Egham Hill, Egham TW20 0EX, U.K.}
\affiliation[b]{Centre for Cosmology, Particle Physics and Phenomenology (CP3), Universit\'{e} Catholique de Louvain, B-1348 Louvain-la-Neuve, Belgium}
\emailAdd{n.kauer@rhul.ac.uk}
\emailAdd{claire.obrien.2012@live.rhul.ac.uk}
\emailAdd{eleni.vryonidou@uclouvain.be}

\abstract{
Signal-background interference effects are studied for $H\to WW$ and $H\to ZZ$ searches in gluon fusion at the LHC.  More specifically, the interference in the channels with semileptonic weak boson pair decay is analysed for light and heavy Higgs masses with minimal and realistic experimental selection cuts.  In the semileptonic decay 
modes, the interference is affected by tree-level background contributions enhanced 
by $1/e^2$ relative to the gluon-fusion continuum background in the fully leptonic decay modes.
We find that for both light and heavy Higgs masses the interference with the loop-induced weak-boson pair background dominates over the interference with the tree-level weak-boson plus jets background for a range of selection cuts.  We therefore 
conclude that 
higher-order background contributions can induce leading interference effects. 
With appropriate background suppression cuts the interference can be reduced to  
the 10\% level for heavy Higgs masses, and to the per mille level for the light SM Higgs.
}


\newcommand{\sla}[1]{\ifmmode%
  \setbox0=\hbox{$#1$}%
  \setbox1=\hbox to\wd0{\hss$/$\hss}\else%
  \setbox0=\hbox{#1}%
  \setbox1=\hbox to\wd0{\hss/\hss}\fi%
  #1\hskip-\wd0\box1 }
  
\newcommand{\calM}{{\cal M}}
\newcommand{\calO}{{\cal O}}

\begin{document}
\maketitle
\flushbottom


\section{Introduction}

In 2012, a scalar resonance at 125 GeV consistent with a Standard Model (SM)-like Higgs boson was discovered at the CERN Large Hadron Collider (LHC) \cite{Aad:2012tfa,Chatrchyan:2012ufa}. Its mass is in the correct range to unitarize the $WW$ scattering, but precision tests of its couplings are necessary in order to determine whether it is the particle predicted by the Higgs mechanism \cite{Higgs:1964ia,Higgs:1964pj,Englert:1964et,Guralnik:1964eu}. Many beyond the Standard Model (BSM) scenarios allow for more scalar particles, and searches for heavier Higgs-like bosons in various channels are in progress.

At Tevatron and LHC energies, gluon fusion is the dominant Higgs boson production mechanism \cite{Georgi:1977gs}. Unfortunately, this process suffers from large higher order corrections and strong scale dependence at leading order (LO), motivating its calculation to higher orders in QCD \cite{Dawson:1990zj,Djouadi:1991tka,Graudenz:1992pv,Spira:1995rr,Harlander:2002wh,Anastasiou:2002yz,Ravindran:2003um,Anastasiou:2015ema}. The combined PDF and scale uncertainties determined by the LHC Higgs Cross Section Working Group are of $\calO(10\%)$ \cite{Dittmaier:2011ti,Dittmaier:2012vm,Heinemeyer:2013tqa}.  Any other effects of the same order must be quantified, and it has been shown that interference effects, particularly at higher Higgs invariant masses, can be of similar size.
In the SM, interference between the Higgs signal and continuum background in 
$gg\ (\to H)\to VV$ ($V=W,Z$) and including fully leptonic decays has been studied in refs.\ \cite{Glover:1988fe,Glover:1988rg,Binoth:2006mf,Campbell:2011cu,Kauer:2012ma,Passarino:2012ri,Kauer:2012hd,Bonvini:2013jha,Kauer:2013qba,Campbell:2013una,Moult:2014pja,Ellis:2014yca,Campanario:2012bh,Campbell:2014gua,Li:2015jva}.\footnote{Predictions for $gg\to\ell\ell\nu\nu+0,1$ jets have been presented in ref.\ \cite{Cascioli:2013gfa}.  SM Higgs-continuum interference in the $H\to VV$ decay modes at a $e^+e^-$
collider has been investigated in ref.\ \cite{Liebler:2015aka}.
}  
Higgs-continuum interference results for a heavy SM Higgs boson have been presented in refs.\ \cite{Campbell:2011cu,Kauer:2012ma,Passarino:2012ri,Campanario:2012bh,Bonvini:2013jha,Kauer:2013qba,Moult:2014pja}.  We note that all 
Higgs-continuum interference calculations are at LO,  except for refs.\ \cite{Bonvini:2013jha,Moult:2014pja,Li:2015jva}, where approximate higher-order corrections have been  
calculated. The technical bottleneck of an unapproximated NLO calculation of $gg\to VV$ and its Higgs-continuum interference is the computation of the 2-loop multiscale integrals needed for the virtual corrections of the continuum background, which is in progress \cite{Caola:2015ila,vonManteuffel:2015msa,Melnikov:2015laa}.  Furthermore, we note that the interfering $gg\to VV$ continuum background at LO is 
formally part of the NNLO corrections to $pp\to VV$ \cite{Cascioli:2014yka,Gehrmann:2014fva}.

For a 125 GeV Higgs boson resonance, the $WW$ semileptonic decay mode has the highest branching fraction of any decay mode with a triggerable lepton and the $ZZ$ semileptonic decay is the third highest \cite{Dittmaier:2011ti,Dittmaier:2012vm,Heinemeyer:2013tqa}. Due to the large $V$+jets background, the semileptonic channels have often been neglected in favour of the fully leptonic ones. But, the semileptonic channels have several kinematic features that allow effective background reduction and given their large rates they merit further study. Both ATLAS \cite{Aad:2012oxa,Aad:2012me,ATLAS:2012mja,Micco:2013vma,Diglio:2014vpa} and CMS \cite{CMS:2013cda,CMS:2015mda,CMS:2015lda,Khachatryan:2015cwa,Pelliccioni:2015hva} have therefore included them in recent studies.  Phenomenological studies have also been carried out (without taking into account interference effects): the semileptonic $ZZ$ channel has been studied for the LHC in ref.\ \cite{Hackstein:2010wk}, and the semileptonic $WW$ channel has been studied for the Tevatron in refs.\ \cite{Dobrescu:2009zf,Lykken:2011uv} and for the LHC in ref.\ \cite{Kao:2012zj}.

An interesting aspect of this particular decay channel is that in addition to the $gg\to VV$ loop continuum,  a tree-level background arises from $ g g \to V q \bar{q}$, with  $V$ decaying leptonically. In this work we will focus on the semileptonic decay mode and for the first time quantify the signal-background interference effects, including both the continuum and tree-level backgrounds. 

This paper is organised as follows: in section \ref{sec:calculation} we review the details of our calculation. In section \ref{sec:results}, we present Higgs signal cross sections and distributions for $gg\to H\to VV$ with semileptonic decay, with minimal and realistic experimental selection cuts taking into account the interference with both the tree- and loop-level backgrounds. In section \ref{sec:conclusions}, we summarize our findings.


\section{Calculational details\label{sec:calculation}}

We consider the hadron-level Higgs signal processes
\begin{align} 
pp &\to H\to W^-W^+\to \ell\bar{\nu}_\ell\, jj\,, \label{hlwm}\\
pp &\to H\to W^+W^-\to \bar{\ell}\nu_\ell\, jj\,, \label{hlwp}\\
pp &\to H\to ZZ\to \ell\bar{\ell}\, jj\,, \label{hlz}
\end{align}
and calculate integrated cross sections and differential distributions for the 
parton-level subprocesses
\begin{align} 
gg &\to H\to W^-W^+\to \ell\bar{\nu}_\ell\, q_u \bar{q}_d \,, \label{plwm} \\
gg &\to H\to W^+W^-\to \bar{\ell}\nu_\ell\, \bar{q}_u q_d \,, \label{plwp}\\
gg &\to H\to ZZ\to \ell\bar{\ell}\, q_u \bar{q}_u \,,  \label{hlzu}\\
gg &\to H\to ZZ\to \ell\bar{\ell}\, q_d \bar{q}_d   \label{hlzd}
\end{align}
including full signal-background interference at LO.
For consistency, the signal cross section is also evaluated at LO.%
\footnote{The $H\gamma\gamma$ effective vertex is not included.}
The hadron-level processes (\ref{hlwm}), (\ref{hlwp}) and (\ref{hlz}) 
also receive contributions from subprocesses of the type 
$g q \to \ell \bar{\nu}_\ell\, g q $ and 
$g q \to \ell \bar{\ell}\, g q $. These crossed subprocesses feature a $t$-channel Higgs progagator.
Using {\sc MadGraph5\_aMC@NLO} \cite{Alwall:2014hca}, we have verified in the heavy top limit that the crossed contributions are several orders of magnitude smaller than the $s$-channel contributions (\ref{plwm}) -- (\ref{hlzd}). 
We therefore neglect the crossed contributions in this study.

\begin{figure}[tb]
\vspace{0.cm}
\centering
\includegraphics[height=2.6cm, clip=true]{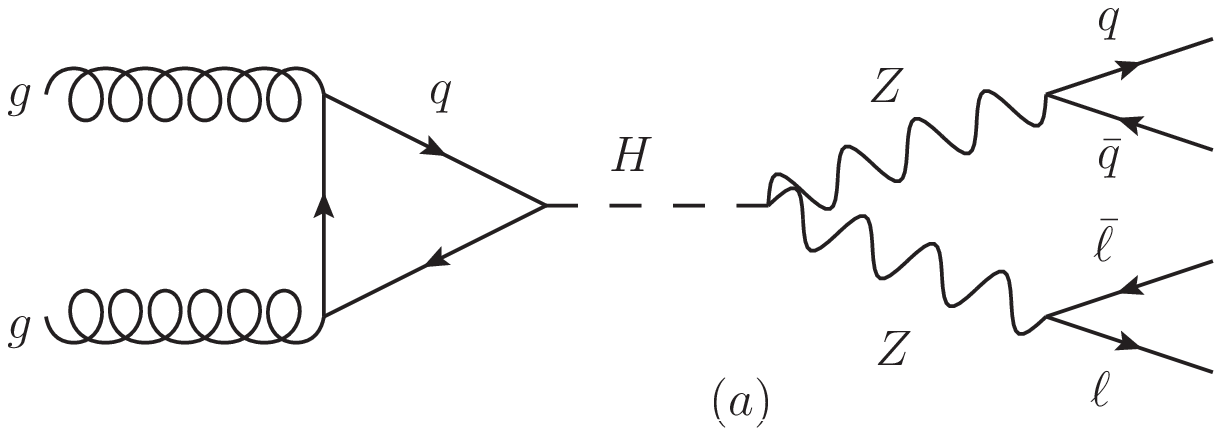}
\includegraphics[height=2.6cm, clip=true]{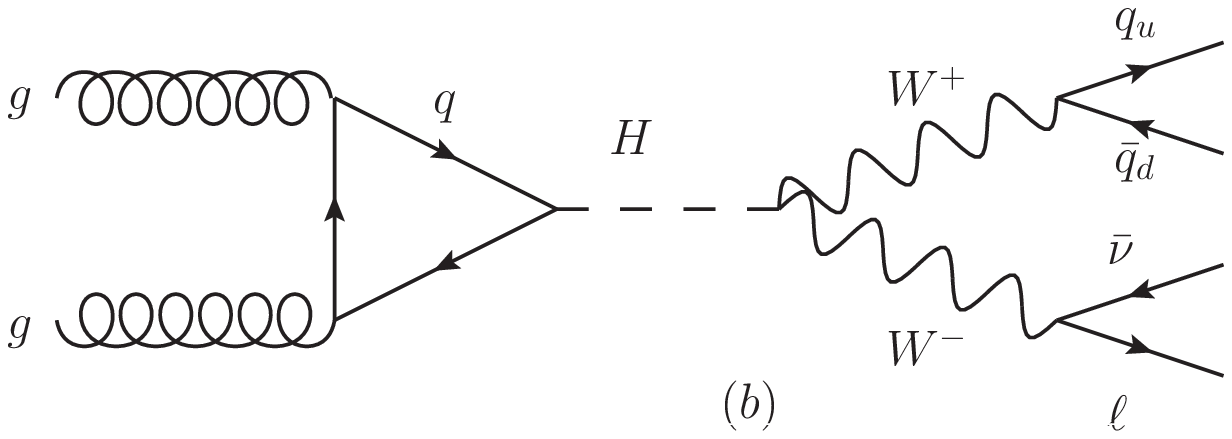}
\caption{\label{fig:sig}
Representative Feynman diagrams for the signal processes considered: (a) $gg\to H \to ZZ \to \ell\bar{\ell}q\bar{q}$ ($q$ can be an up- or down-type quark, $q_{u,d}$) and (b) $gg\to H \to WW \to \ell\bar{\nu}_\ell\bar{q}_dq_u$ (the charge-conjugated process is also considered).}
\end{figure}

\begin{figure}[tb]
\vspace{0.cm}
\centering
\includegraphics[height=2.6cm, clip=true]{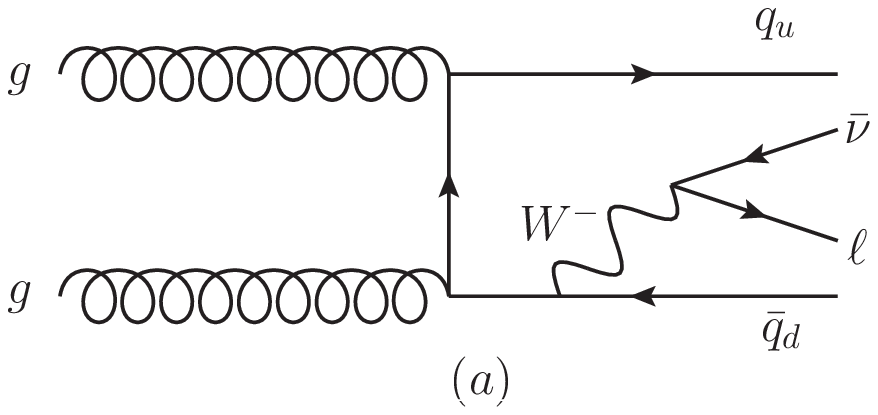}
\includegraphics[height=2.6cm, clip=true]{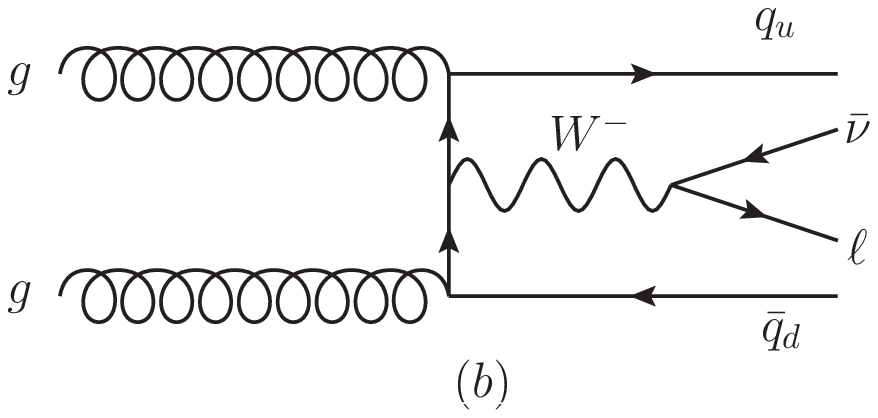}
\includegraphics[height=2.6cm, clip=true]{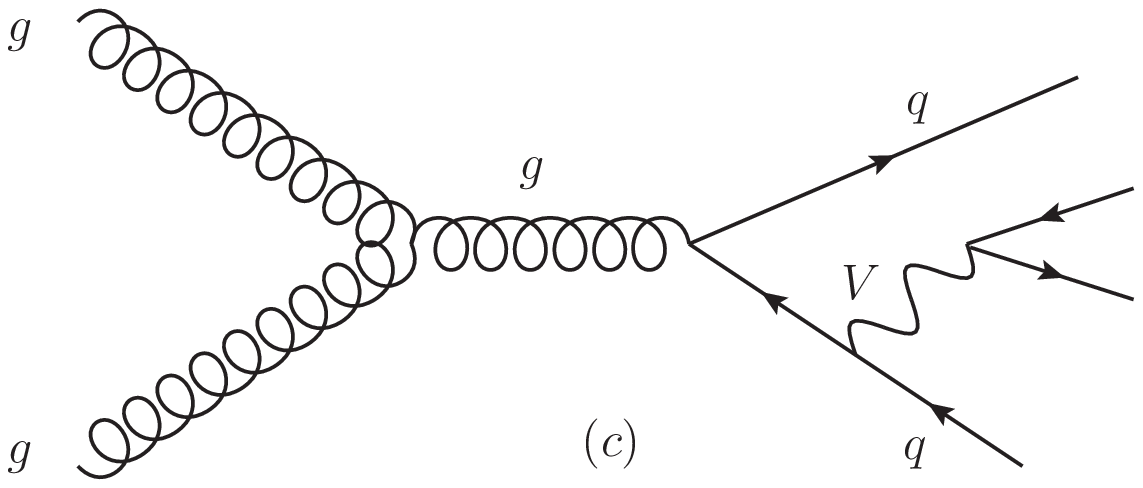}
\caption{\label{fig:tree}
Representative tree-level background diagrams of $\calO(g_s^2 e^2)$ that interfere with the signal diagrams in figure \ref{fig:sig}.}
\end{figure}

\begin{figure}[tb]
\vspace{0.cm}
\centering
\includegraphics[height=2.6cm, clip=true]{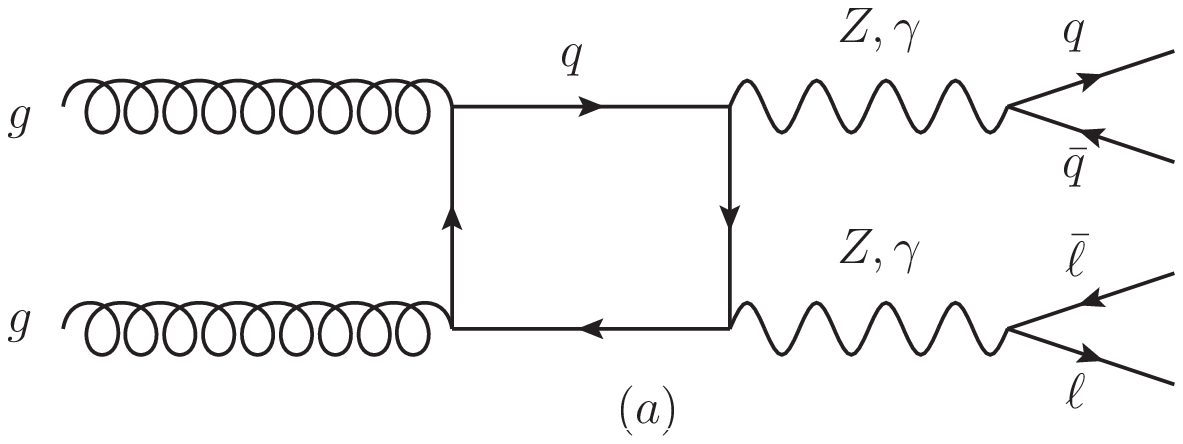}
\includegraphics[height=2.6cm, clip=true]{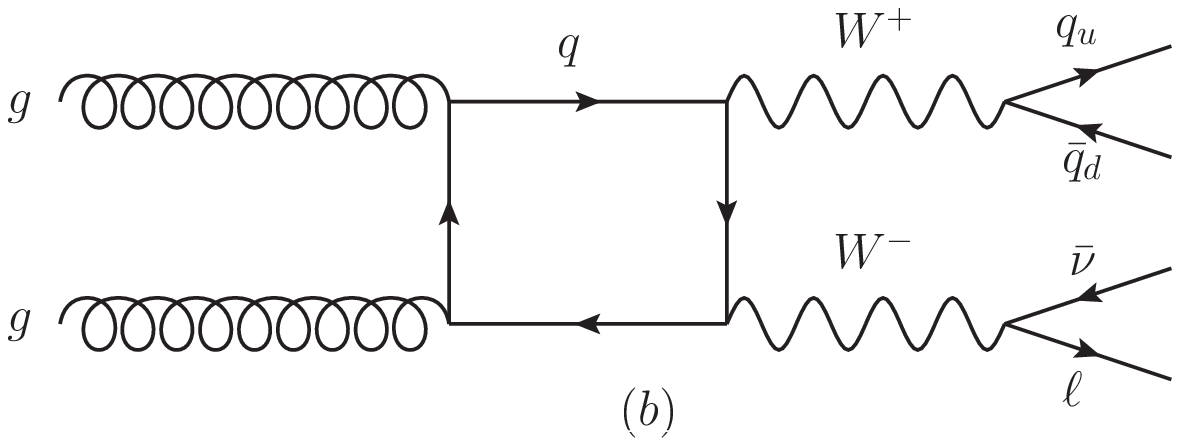}
\caption{\label{fig:cont} Representative one-loop background diagrams of $\calO(g_s^2 e^4)$ that interfere with the signal diagrams in figure \ref{fig:sig}.}
\end{figure}

\begin{figure}[tb]
\vspace{0.cm}
\centering
\includegraphics[height=2.6cm, clip=true]{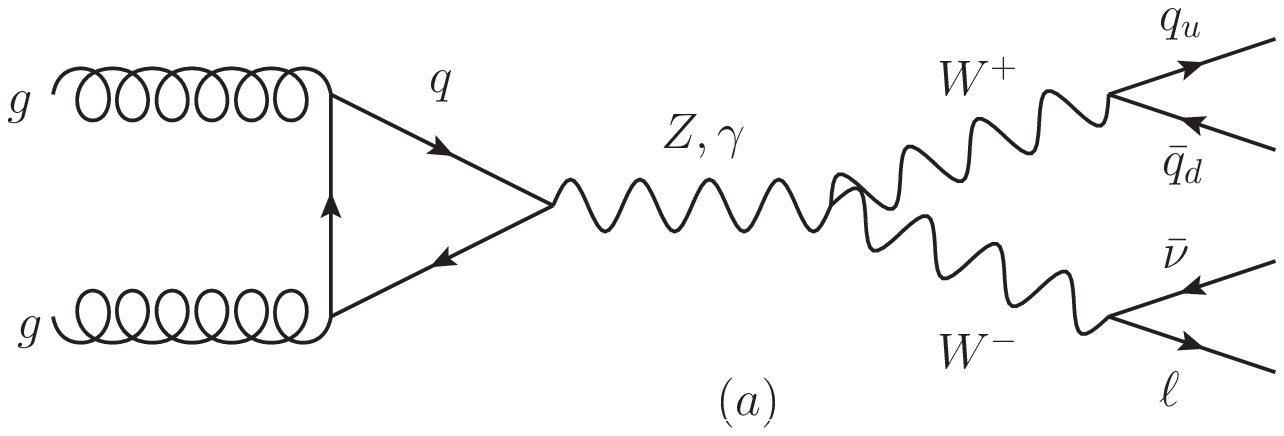}
\includegraphics[height=2.6cm, clip=true]{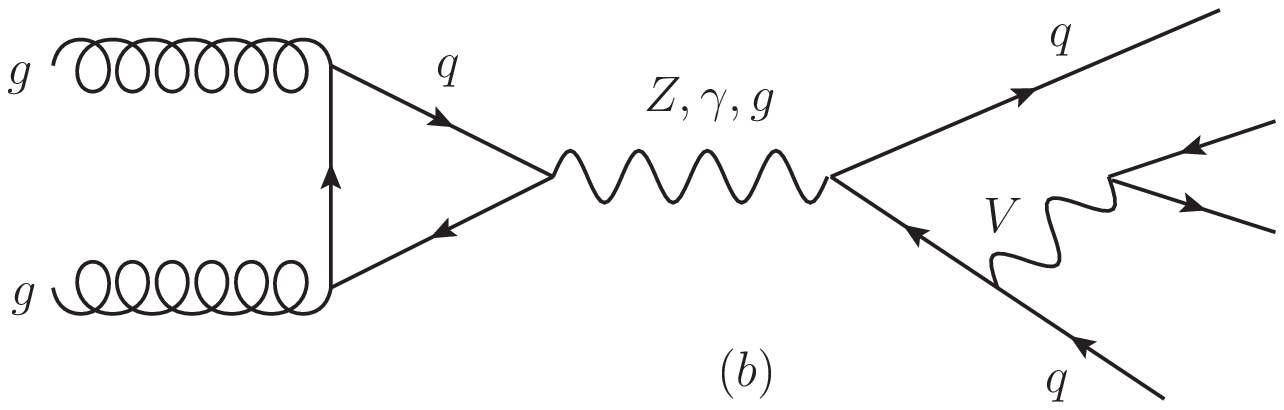}
\caption{\label{fig:triangles}
Representative triangle diagrams that formally contribute (see main text).}
\end{figure}

Figures \ref{fig:sig}--\ref{fig:triangles} show representative Feynman diagrams for the different amplitude contributions. The amplitude $\calM$ is decomposed as
follows:
\begin{align} 
\calM &= \calM_{signal} + \calM_{background}\,,\\
\calM_{background} &= \calM_{tree} + \calM_{loop}\,,
\end{align}
where $\calM_{tree}$ contains all tree-level contributions and $\calM_{loop}$ contains all quark-loop contributions.%
\footnote{We note that the interference between the tree-level and loop background contributions is at the 1\% level or less.  This was verified for the $ZZ$ channels and all cut sets using \textsf{gg2VV}.}
We introduce the following notation for amplitude contributions to 
integrated cross sections:
\begin{align} 
S &\sim \left|\calM_{signal}\right|^2\\
I_{tree} &\sim 2\,\mathrm{Re}(\calM_{signal}^\ast\,\calM_{tree})\\
I_{loop} &\sim 2\,\mathrm{Re}(\calM_{signal}^\ast\,\calM_{loop})\\
I_{full} &\sim 2\,\mathrm{Re}(\calM_{signal}^\ast\,\calM_{background})
\end{align}
In addition to the signal contribution $S$ (see figure \ref{fig:sig}), we include $I_{tree}$, i.e.\ the interference with the LO (tree-level) background diagrams, which are of $\calO(g_s^2 e^2)$ (see figure \ref{fig:tree}). 
For $M_{VV} > 2M_V$, the interference between the $gg\to H\to VV$ 
signal process and
the $gg\to VV$ continuum background is known to be large.  
$I_{loop}$, i.e.\ the interference with the 
$gg\to (V\to \text{leptons})(V\to \text{quarks})$ 
continuum amplitude, which is of $\calO(g_s^2 e^4)$ (see figure \ref{fig:cont}), is therefore also taken 
into account. The sum of $I_{tree}$ and $I_{loop}$ is denoted by $I_{full}$.
The $gg\to VV$ process is loop induced and the complete amplitude is 
UV and IR finite since no $ggVV$ counter term or real corrections to a tree-level
Born term exist.
As in the fully leptonic case, in addition to the box diagrams 
shown in figure \ref{fig:cont}, triangle diagrams formally
also contribute at the same order (see figure \ref{fig:triangles}).
Since we consider $M_{VV} < 2M_V$, the weak boson pair is not treated in the 
narrow-width approximation, and the singly-resonant triangle diagrams 
shown on the right-hand side of figure \ref{fig:triangles} have to be considered.
The triangle diagrams with intermediate photon or gluon vanish due to 
Furry's theorem \cite{Furry:1937zz}.
The $Z$ boson vector coupling contribution vanishes for the same reason.  
Regarding the $Z$ boson axial-vector coupling contribution: in the case of 
four massless final state leptons, the sum of all triangle diagrams has analytically 
been shown to vanish for $V=W$ \cite{Binoth:2006mf} (and $V=Z$ if $m_q=0$ in the 
loop \cite{Campbell:2011bn}).  For $V=W$ this result carries over to the 
semileptonic decay mode, because the $Wf\bar{f}$ coupling is flavour independent 
($V_{CKM}=\mathbf{1}$).  For $V=Z$ and semileptonic decay with finite $M_{t,b}$, 
we have checked numerically that the contribution of the triangle diagrams is 
consistent with zero.\footnote{The complete triangle amplitude contribution is 
nevertheless included in our calculation.}

$\calM_{loop}$ is of $\calO(g_s^2 e^4)$ and hence of the 
same order as the $\calO(e^2)$ virtual electroweak (EW) corrections to 
$\calM_{tree}$.  
The complete set of these virtual corrections will not be finite 
and is hence not taken into account.  In particular all self-energy corrections and 
all diagrams with boson propagators in the loop are not included.  We argue that
these are part of the next-to-leading order (NLO) EW corrections to $I_{tree}$ and are genuinely suppressed
by $\calO(\alpha)$.  Given that $I_{tree}$ at LO yields a small (tiny) 
correction to $I_{loop}$ for integrated results when LHC (background suppression) 
cuts are applied (see section \ref{sec:results}) this treatment is justified.  We argue similarly
that neglecting the NLO QCD corrections to $I_{tree}$ in our calculation is
justified.

To obtain and independently cross check the results presented in this work, we follow two independent approaches. In the first, we implement the amplitudes in the publicly available program \textsf{gg2VV} \cite{gg2VV}, while in the second we make use of the also public automated {\sc MadGraph5\_aMC@NLO} framework \cite{Alwall:2014hca}.  

For the computation in \textsf{gg2VV}, the amplitudes are implemented and calculated using \textsf{FeynArts} \cite{Hahn:2000kx}, \textsf{Formcalc} \cite{Hahn:1998yk} and \textsf{LoopTools} \cite{Hahn:1998yk}, with code adaptation for compatibility with \textsf{gg2VV}.  A fixed-width 
Breit-Wigner propagator is employed for the weak bosons and the Higgs boson. 
The width parameter $\gamma_H$ of the complex pole of the Higgs propagator, 
as defined in eq.\ (16) of ref.\ \cite{Goria:2011wa}, is calculated using 
the \textsc{HTO} code \cite{Goria:2011wa}.
The box diagrams shown in figure \ref{fig:cont}
are affected by numerical instabilities when Gram determinants 
approach zero.  In these critical phase space regions the amplitude 
is evaluated in quadruple precision in \textsf{gg2VV}, and residual instabilities 
are eliminated by requiring that $p_{T,W}$ and $p_{T,Z}$ are larger than $1$ GeV.
This criterion is also applied to amplitudes which are 
not affected by numerical instabilities, in order to obtain consistent 
cross section-level results.  The numerical effect of this technical cut
has been shown to be small \cite{Kauer:2012ma,Campbell:2013una}.
The diagrams in figures \ref{fig:sig}--\ref{fig:cont} show the different kinematical structures that appear in the various amplitude components.
In addition to the Higgs and weak boson resonance peaks, the tree level diagrams exhibit mass singularities that must be dealt with when integrating the phase space.  In \textsf{gg2VV} this is achieved with the multi-channel Monte Carlo integration 
technique \cite{Berends:1994pv}, in which every kinematic structure 
has its own mapping from random variables to the phase space configuration 
such that peaks or singularities in the amplitude are compensated, and the 
inverse Jacobi determinants of all mappings are summed to give the inverse weight at
each phase space point. The bottom diagram in figure \ref{fig:tree} does not require its own mapping because the $s$-channel singularity coincides with a vanishing phase space volume.  Additional details of the phase space implementation and validation 
in \textsf{gg2VV} can be found in ref.\ \cite{Kauer:2015hia}.

The computation within {\sc MadGraph5\_aMC@NLO} makes use of the recent development of the automation of event generation for loop-induced processes \cite{Hirschi:2015iia,launchpad}, as well as the optimisations made in relation to large-scale Monte-Carlo production \cite{large-scale}, and in particular the new interference module. The one-loop amplitudes in {\sc MadGraph5\_aMC@NLO} are obtained with the help of {\sc MadLoop} \cite{Hirschi:2011pa}, which computes one--loop matrix elements using the {\sc OPP} integrand--reduction method~\cite{Ossola:2006us} (as implemented in {\sc CutTools}~\cite{Ossola:2007ax}). 

Cross checking the results is important, in particular because the \textsf{gg2VV} code implements considerable changes to how the phase space integration is performed.
We therefore verified all results presented here with independent calculations using \textsf{gg2VV} and {\sc MadGraph5\_aMC@NLO}. The agreement between the two codes is excellent: an example for the process $gg \to H\to W^-W^+\to \ell\bar{\nu}_\ell\, q_u \bar{q}_d$ is given in table \ref{tab:mg5} in section \ref{sec:results}, and other results show similar agreement.  In addition to this cross section level validation, we also successfully compared the spin/polarisation-summed mod-squared amplitudes for all processes at two phase space points and 
found 4-significant-digit agreement for all $S$, $I_{tree}$, $I_{loop}$ and $I_{full}$.


\section{Results\label{sec:results}}

In this section, we present integrated cross sections and differential distributions 
for the considered Higgs signal processes taking into account the interference 
with the tree- and loop-level background contributions. 
We do not give results for process (\ref{plwp})
with $\bar{\ell}\nu_\ell\, \bar{q}_u q_d$ final state, because
they are identical to the results for process 
(\ref{plwm}) with $\ell\bar{\nu}_\ell\, q_u \bar{q}_d$
final state due to the $CP$ symmetry of the amplitude 
(and the symmetry of the applied selection cuts).\footnote{
To validate our calculations, we have numerically verified explicitly the agreement 
of integrated and differential results for processes 
(\ref{plwm}) and (\ref{plwp}).}
To obtain numerical results, the renormalisation and factorisation scales are set to $M_{\ell\bar{\nu}q\bar{q}}/2$ for $V=W$ and $M_{\ell\bar{\ell}q\bar{q}}/2$ for $V=Z$. The MSTW2008LO \cite{Martin:2009iq} PDF set is used with default $\alpha_s$. 
The CKM matrix is approximated by the unit matrix, which causes a negligible error \cite{Kauer:2012ma}.
As input parameters, we use the recommendation of the 
LHC Higgs Cross Section Working Group in App.~A of ref.\ \cite{Dittmaier:2011ti}
with $G_\mu$ scheme and LO weak boson widths for consistency.
More specifically, $M_W = 80.398$ GeV, $M_Z = 91.1876$ GeV, $\Gamma_W = 2.141$ GeV, $
\Gamma_Z = 2.4952$ GeV, $M_t = 172.5$ GeV, $M_b = 4.75$ GeV, $G_F = 1.16637\cdot 10^{-5}$ GeV$^{-2}$ are used. The Higgs width parameter $\gamma_H$ (see section \ref{sec:calculation}) is set to 
4.098973 MeV and 26.59768 GeV for a Higgs mass of 125.5 and 400 GeV, respectively.
Finite top and bottom quark mass effects 
are included.  Lepton and light quark masses are neglected.
Proton-proton collision energies of 8, 13 and 14 TeV are considered.

Results for all processes are computed for the following three sets of cuts:\footnote{No jet clustering algorithm is applied.}\\[.1cm]
$\bullet$ \textit{minimal cuts}:  $p_{Tj} > 4$ GeV, and $M_{Z/\gamma}>4$ GeV to eliminate soft photon singularities\\[.15cm]
$\bullet$ \textit{LHC cuts} (mainly detector resolution): minimal cuts and $p_{T\ell} > 20$ GeV, $|\eta_\ell| < 2.5$,\\
\phantom{$\bullet$} $p_{Tj} > 25$ GeV, $|\eta_j| < 4.5$, and for $H\to WW$ in addition: $\sla{p}_T > 20$ GeV\\[.15cm] 
$\bullet$ \textit{background suppression cuts} for a $400$ GeV SM Higgs boson \cite{Aad:2012me}: LHC cuts and\\
\phantom{$\bullet$} $|M_{jj} - M_V| < 5\,\Gamma_V$, 
$p_{Tj,1st} > 60$ GeV, $p_{Tj,2nd} > 40$ GeV, $|\eta_j| < 2.8$, $\Delta R_{jj} < 1.3$
\\[.15cm]

For the processes with intermediate $W$-boson pair, we also calculate results using the \textit{background suppression cuts} proposed in ref.\ \cite{Kao:2012zj} for a $125.5$ GeV Higgs boson at $\sqrt{s} \approx 14$ TeV:\\[.1cm]
$\bullet$ $p_{Tj,1st} > 30$ GeV, $p_{Tj,2nd} > 20$ GeV, $65$ GeV $< M_{jj} < 95$ GeV, $p_{T\ell} < 30$ GeV, $\sla{p}_T < 40$ GeV, $|\eta_j|<5$, $|\eta_\ell|<2.5$, $M_{\ell\nu} < 45$ GeV, $M_{jj\ell\nu} < 130$ GeV, $\Delta R_{j\ell}>0.2$\\[.15cm]

In tables \ref{tab:WWqqlv}--\ref{tab:WWqqlvsmbkg}, integrated cross sections are given for the $WW$ process for the cuts specified above, while in tables \ref{tab:ZZlluu}--\ref{tab:ZZlldd13} the results for $ZZ$ are shown. To illustrate the relative effect of the 
signal-background interference, the ratios $R_i = (S + I_i)/S$ are also displayed. The results presented in the tables and plots have been obtained with \textsf{gg2VV} and cross checked with  {\sc MadGraph5\_aMC@NLO}.

 
\begin{table}[tb]
\vspace*{0.cm}
\centering
{\footnotesize
\renewcommand{\arraystretch}{1.2}
\begin{tabular}{|c|l|c|ccc|ccc|}
\cline{1-3}
\multicolumn{3}{|c|}{$gg\to H\to W^-W^+\to \ell\bar{\nu}_\ell q_u \bar{q}_d$} & \multicolumn{6}{|c}{} \\ 
\cline{4-9}
  \multicolumn{3}{|c|}{$\sigma$ [fb], $pp$, $\sqrt{s}=8$ TeV} &
\multicolumn{3}{c|}{interference} & \multicolumn{3}{c|}{ratio} \\ 
\hline
$M_H$ [GeV] & cuts  & $S$ & $I_{tree}$    & $I_{loop}$ & $I_{full}$  &$R_{tree}$ & $R_{loop}$ & $R_{full}$ \\ \hline

$125.5$ &  min.      &       67.28(9)        &       -2.47(2)        &       -4.99(1)        &       -7.48(9)        &       0.963(2)        &       0.926(2)        &       0.889(3)        \\ \hline
$125.5$ & LHC   &       1.978(6)        &       0.266(4)        &       -2.647(6)       &       -2.38(3)        &       1.135(5)        &       -0.338(4)       &       -0.20(2)        \\ \hline

$400$ & bkg.  &       13.30(2)        &       -0.0054(2)      &       -1.052(5)       &       -1.058(4)       &       1.000(2)        &       0.921(2)        &       0.920(2)        \\ \hline
\end{tabular}}
\caption{\label{tab:WWqqlv}Cross sections for the signal process $gg\to H\to W^-W^+\to \ell\bar{\nu}_\ell q_u \bar{q}_d$ ($S$) and its interference with the tree-level ($I_{tree}$) and quark-loop ($I_{loop}$) $gg$ background contributions as well 
as $I_{full}=I_{tree}+I_{loop}$ in $pp$ collisions at $\sqrt{s}=8$ 
TeV with minimal and LHC cuts for a $125.5$ GeV SM Higgs boson and background suppression 
cuts for a $400$ GeV SM Higgs boson (see main text). To illustrate the relative effect of the 
signal-background interference, the ratios $R_i = (S + I_i)/S$ are given.
Cross sections are given for single lepton and quark flavour combinations.
The integration error is displayed in brackets.}
\end{table}


\begin{table}[tb]
\vspace*{0.cm}
\centering
{\footnotesize
\renewcommand{\arraystretch}{1.2}
\begin{tabular}{|c|l|c|ccc|ccc|}
\cline{1-3}
\multicolumn{3}{|c|}{$gg\to H\to W^-W^+\to \ell\bar{\nu}_\ell q_u \bar{q}_d$} & \multicolumn{6}{|c}{} \\ 
\cline{4-9}
  \multicolumn{3}{|c|}{$\sigma$ [fb], $pp$, $\sqrt{s}=13$ TeV} &
\multicolumn{3}{c|}{interference} & \multicolumn{3}{c|}{ratio} \\ 
\hline
$M_H$ [GeV] & cuts  & $S$ & $I_{tree}$    & $I_{loop}$ & $I_{full}$  &$R_{tree}$ & $R_{loop}$ & $R_{full}$ \\ \hline

125.5 & min.      &      162.1(3)        &      -5.9(1)        &       -15.36(4)        &       -21.2(4)        &       0.964(3)        &       0.905(2)        &       0.869(3)        \\ \hline
125.5 & LHC   &       5.56(2)        &       0.83(3)        &       -8.34(3)       &       -7.51(7)        &       1.15(2)        &       -0.500(5)       &       -0.35(2)        \\ \hline
400 & bkg.  &       43.10(4)       &       -0.018(2)      &       -4.29(2)       &       -4.30(4)       &       1.000(2)        &       0.901(2)        &       0.900(2)        \\ \hline
\end{tabular}}
\caption{\label{tab:WWqqlv13}Cross sections for the signal process $gg\to H\to W^-W^+\to \ell\bar{\nu}_\ell q_u \bar{q}_d$ and its interference with the tree-level and quark-loop $gg$ background contributions in $pp$ collisions at $\sqrt{s}=13$ TeV. 
Other details as in table \ref{tab:WWqqlv}.}
\end{table}


\begin{table}[tb]
\vspace*{0.cm}
\centering
{\footnotesize
\renewcommand{\arraystretch}{1.2}
\begin{tabular}{|c|c|ccc|ccc|}
\cline{1-2}
\multicolumn{2}{|c|}{$gg\to H\to W^-W^+\to \ell\bar{\nu}_\ell q_u \bar{q}_d$} & \multicolumn{3}{|c}{} \\ 
\multicolumn{2}{|c|}{$\sigma$ [fb], $pp$, $M_H=125.5$ GeV} & \multicolumn{3}{|c}{} \\ 
\cline{3-8}
  \multicolumn{2}{|c|}{background suppression cuts} &
\multicolumn{3}{c|}{interference} & \multicolumn{3}{c|}{ratio} \\ 
\hline
$\sqrt{s}$ [TeV] & $S$ & $I_{tree}$    & $I_{loop}$ & $I_{full}$  &$R_{tree}$ & $R_{loop}$ & $R_{full}$ \\ \hline
13  &       42.16(5)       &       -0.0148(5)      &       0.0264(2)      &       0.0118(6)       &       1.000(2)        &       1.001(2)        &       1.000(2)        \\ \hline
14  &        47.44(5)        &       -0.0164(5)      &       0.029(1)        &       0.0131(6) &       1.000(2)        &       1.001(2)        &       1.000(2)        \\ \hline
\end{tabular}}
\caption{\label{tab:WWqqlvsmbkg}Cross sections for the signal process $gg\to H\to W^-W^+\to \ell\bar{\nu}_\ell q_u \bar{q}_d$ and its interference with the tree-level and quark-loop $gg$ background contributions in $pp$ collisions at $13$ and $14$ TeV with background suppression cuts for a $125.5$ GeV SM Higgs boson at $\sqrt{s}\approx 14$ TeV (see main text).  Other details as in table \ref{tab:WWqqlv}.}
\end{table}
 
 
\begin{table}[tb]
\vspace*{0.cm}
\centering
{\footnotesize
\renewcommand{\arraystretch}{1.2}
\begin{tabular}{|c|l|c|ccc|ccc|}
\cline{1-3}
\multicolumn{3}{|c|}{$gg\to H\to ZZ\to \ell\bar{\ell} q_u \bar{q}_u$} & \multicolumn{6}{|c}{} \\ 
\cline{4-9}
  \multicolumn{3}{|c|}{$\sigma$ [fb], $pp$, $\sqrt{s}=8$ TeV} &
\multicolumn{3}{c|}{interference} & \multicolumn{3}{c|}{ratio} \\ 
\hline
$M_H$ [GeV] & cuts  & $S$ & $I_{tree}$    & $I_{loop}$ & $I_{full}$  &$R_{tree}$ & $R_{loop}$ & $R_{full}$ \\ \hline
$125.5$ &  min.        &       1.954(2)        &       -0.19(2)        &       -0.3442(6)      &       -0.535(9)       &       0.902(7)        &       0.824(2)        &       0.726(5)        \\ \hline
$125.5$ &  LHC     &       0.1164(7)       &       0.0173(9)       &       -0.1940(4)      &       -0.177(2)       &       1.15(2) &       -0.667(7)       &       -0.52(2)
        \\ \hline
$400$ &  bkg.  &       1.256(2)        &       -0.00082(4)     &       -0.0908(3)      &       -0.0917(3)      &       0.999(2)        &       0.928(2)        &       0.927(2)        \\ \hline
\end{tabular}}
\caption{\label{tab:ZZlluu}Cross sections for the signal process $gg\to H\to ZZ\to \ell\bar{\ell} q_u \bar{q}_u$ and its interference with the tree-level and quark-loop $gg$ background contributions in $pp$ collisions at $\sqrt{s}=8$ TeV. $\gamma^\ast$ background contributions are included. Other details as in table \ref{tab:WWqqlv}.} 
\end{table}
  

\begin{table}[tb]
\vspace*{0.cm}
\centering
{\footnotesize
\renewcommand{\arraystretch}{1.2}
\begin{tabular}{|c|l|c|ccc|ccc|}
\cline{1-3}
\multicolumn{3}{|c|}{$gg\to H\to ZZ\to \ell\bar{\ell} q_u \bar{q}_u$} & \multicolumn{6}{|c}{} \\ 
\cline{4-9}
  \multicolumn{3}{|c|}{$\sigma$ [fb], $pp$, $\sqrt{s}=13$ TeV} &
\multicolumn{3}{c|}{interference} & \multicolumn{3}{c|}{ratio} \\ 
\hline
$M_H$ [GeV] & cuts  & $S$ & $I_{tree}$    & $I_{loop}$ & $I_{full}$  &$R_{tree}$ & $R_{loop}$ & $R_{full}$ \\ \hline
125.5 & min.        &       4.79(4)        &       -0.45(3)        &       -1.088(2)      &       -1.54(3)       &       0.91(2)        &       0.773(9)        &       0.68(1)        \\ \hline
125.5 & LHC     &       0.375(2)       &       0.063(7)       &       -0.612(1)      &       -0.552(6)       &       1.17(2)       &       -0.633(6)       &       -0.47(2)
        \\ \hline
400 & bkg.  &       4.043(4)        &      -0.0027(3)     &       -0.3569(9)      &       -0.359(3)       &       0.999(2)        &       0.912(2)        &       0.911(2)        \\ \hline
\end{tabular}}
\caption{\label{tab:ZZlluu13}
Cross sections for the signal process $gg\to H\to ZZ\to \ell\bar{\ell} q_u \bar{q}_u$ and its interference with the tree-level and quark-loop $gg$ background contributions in $pp$ collisions at $\sqrt{s}=13$ TeV. 
Other details as in table \ref{tab:ZZlluu}.} 
\end{table}

  
\begin{table}[tb]
\vspace*{0.cm}
\centering
{\footnotesize
\renewcommand{\arraystretch}{1.2}
\begin{tabular}{|c|l|c|ccc|ccc|}
\cline{1-3}
\multicolumn{3}{|c|}{$gg\to H\to ZZ\to \ell\bar{\ell} q_d \bar{q}_d$} & \multicolumn{6}{|c}{} \\ 
\cline{4-9}
  \multicolumn{3}{|c|}{$\sigma$ [fb], $pp$, $\sqrt{s}=8$ TeV} &
\multicolumn{3}{c|}{interference} & \multicolumn{3}{c|}{ratio} \\ 
\hline
$M_H$ [GeV] &  cuts  & $S$ & $I_{tree}$    & $I_{loop}$ & $I_{full}$  &$R_{tree}$ & $R_{loop}$ & $R_{full}$ \\ \hline
$125.5$ &  min.   &       2.505(4)        &       -0.244(3)       &       -0.443(1)       &       -0.686(6)       &       0.903(2)        &       0.823(2)        &       0.726(3)        \\ \hline
$125.5$ &  LHC   &       0.1498(4)       &       0.022(2)        &       -0.2493(5)      &       -0.227(2)       &       1.146(9)        &       -0.664(5)       &       -0.52(2)        \\ \hline
$400$ &  bkg.  &       1.611(2)        &       -0.00110(4)     &       -0.1167(3)      &       -0.1176(4)      &       0.999(2)        &       0.928(2)        &       0.927(2)        \\ \hline
\end{tabular}}
\caption{\label{tab:ZZlldd}Cross sections for the signal process $gg\to H\to ZZ\to \ell\bar{\ell} q_d \bar{q}_d$ and its interference with the tree-level and quark-loop $gg$ background contributions in $pp$ collisions at $\sqrt{s}=8$ TeV.  Other details as in table \ref{tab:ZZlluu}.}
\end{table}


\begin{table}[tb]
\vspace*{0.cm}
\centering
{\footnotesize
\renewcommand{\arraystretch}{1.2}
\begin{tabular}{|c|l|c|ccc|ccc|}
\cline{1-3}
\multicolumn{3}{|c|}{$gg\to H\to ZZ\to \ell\bar{\ell} q_d \bar{q}_d$} & \multicolumn{6}{|c}{} \\ 
\cline{4-9}
  \multicolumn{3}{|c|}{$\sigma$ [fb], $pp$, $\sqrt{s}=13$ TeV} &
\multicolumn{3}{c|}{interference} & \multicolumn{3}{c|}{ratio} \\ 
\hline
$M_H$ [GeV] &  cuts  & $S$ & $I_{tree}$    & $I_{loop}$ & $I_{full}$  &$R_{tree}$ & $R_{loop}$ & $R_{full}$ \\ \hline
125.5 & min.   &       6.16(2)        &       -0.57(3)       &       -1.396(3)       &       -1.97(2)       &       0.907(5)        &       0.773(3)        &       0.680(4)        \\ \hline
125.5 & LHC   &       0.4809(9)       &       0.077(8)        &       -0.786(2)      &       -0.708(5)       &       1.16(2)        &       -0.635(4)       &       -0.47(2)        \\ \hline
400 & bkg.  &       5.185(5)        &       -0.0038(4)     &       -0.457(1)      &       -0.461(2)      &       0.999(2)        &       0.912(2)        &       0.911(2)        \\ \hline
\end{tabular}}
\caption{\label{tab:ZZlldd13}
Cross sections for the signal process $gg\to H\to ZZ\to \ell\bar{\ell} q_d \bar{q}_d$ and its interference with the tree-level and quark-loop $gg$ background contributions in $pp$ collisions at $\sqrt{s}=13$ TeV. 
Other details as in table \ref{tab:ZZlluu}.}
\end{table}


\begin{table}[tb]
\centering
\vspace*{0.cm}
\centering
{\footnotesize
\renewcommand{\arraystretch}{1.2}	
\begin{tabular}{|l|l|l|l|l|}
\cline{2-4}
\multicolumn{1}{c|}{} & \multicolumn{3}{|c|}{$gg \to H\to W^-W^+\to \ell\bar{\nu}_\ell\, q_u \bar{q}_d$} \\ 
\cline{1-4}
 cuts  &  $S$ & $I_{tree}$  & $I_{loop}$ \\ \hline
 min. (\textsf{gg2VV})     &         67.28(9)   &       -2.47(2)        &       -4.99(1)    \\ \hline
 min. (\textsc{MG5$\_$aMC@NLO}) & 67.19(6) & -2.49(2) & -5.004(3)   \\ \hline 
LHC (\textsf{gg2VV}) &           1.978(6)        &       0.266(4)        &       -2.647(6)       \\ \hline
LHC (\textsc{MG5$\_$aMC@NLO}) & 1.963(3) & 0.264(4) & -2.646(7)   \\ \hline 
bkg. (\textsf{gg2VV}) &         13.30(2)        &       -0.0054(2)      &       -1.052(5)      \\ \hline
bkg. (\textsc{MG5$\_$aMC@NLO}) & 13.30(2) & -0.0057(5) & -1.08(2)   \\ \hline 
\end{tabular}
\caption{\label{tab:mg5} Comparison of cross sections calculated with \textsf{gg2VV} and \textsc{MG5\_aMC@NLO} ($\sqrt{s}=8$ TeV). The \textsf{gg2VV} results are taken from table \ref{tab:WWqqlv}.}}
\end{table}


We notice that for both $WW$ and $ZZ$ the total contribution of the interference strongly depends on the choice of selection cuts. A general observation is that for all sets of cuts and final states the loop background is dominant compared to the tree one. In more detail, for minimal cuts for the light Higgs the interference for $WW$ is at the 5--10\% level and destructive for both the tree and loop backgrounds, resulting in a total reduction of the signal of $\mathcal{O}(10\%)$. For $ZZ$, we notice that for minimal cuts this rises to $\mathcal{O}(30\%)$. With LHC cuts, the loop and total interference become larger than the signal for both final states, and the total interference is heavily dominated by the loop contribution. In the heavy Higgs case, the background suppression cuts force the interference to fall to 7--10\%. In this case, the interference with the tree level background is completely negligible. 

Finally, for the light Higgs mass with appropriate background suppression cuts \cite{Kao:2012zj} the interference for $WW$ is reduced to the sub-percent level as the invariant mass is forced to remain very close to the mass of the Higgs, removing both the tree- and loop-level backgrounds. We expect a similar behaviour for $ZZ$ with an appropriate set of cuts. 

To illustrate our validation, in table \ref{tab:mg5} we present a detailed comparison of integrated results for the signal, tree and loop interference calculated with \textsf{gg2VV} and {\sc MadGraph5\_aMC@NLO} for the $WW$ process.  All results agree within the integration errors. Similar agreement is achieved for all other channels at the integrated level, but these comparisons are not shown here for brevity.

\begin{figure}[tb]
\centering
\includegraphics[width = 0.9\textwidth]{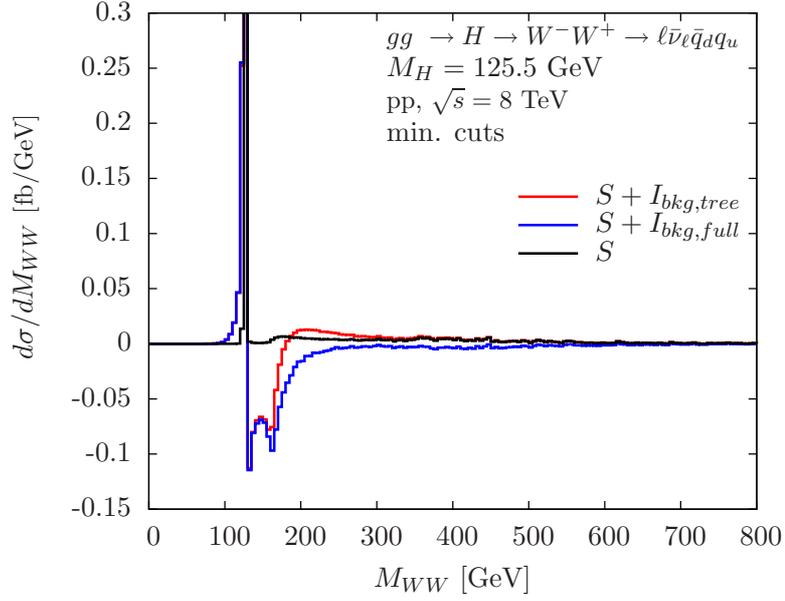}\hfill\hspace{3mm}
\caption{\label{fig:WWqqlvmin} Invariant $WW$ mass distributions for the signal 
process $gg\to H\to W^-W^+\to \ell\bar{\nu}_\ell q_u \bar{q}_d$ ($S$)
and including its interference with the tree-level ($S+I_{bkg,tree}$) and 
in addition quark-loop ($S+I_{bkg,full}$) $gg$ background contributions
in $pp$ collisions at $\sqrt{s}=8$ TeV for a $125.5$ GeV SM Higgs boson.
Minimal cuts are applied (see main text).  Other details as in table \ref{tab:WWqqlv}.}
\end{figure}

\begin{figure}[tb]
\centering
\includegraphics[width = 0.9\textwidth]{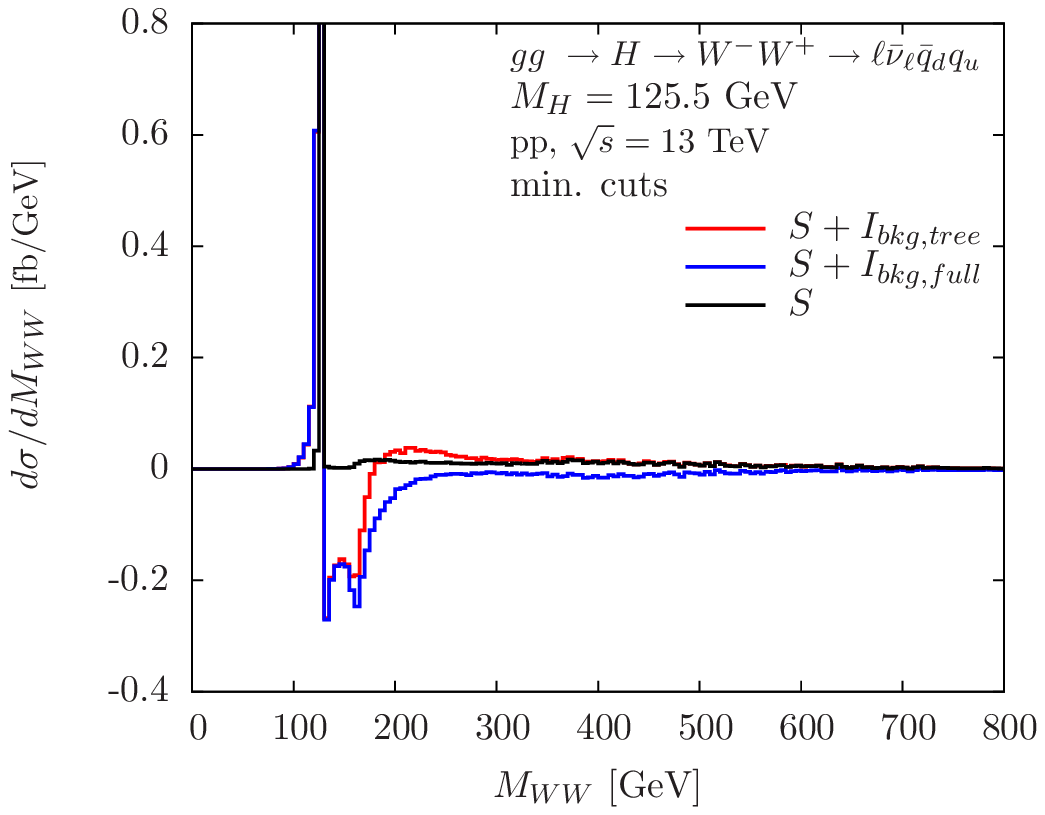}\hfill\hspace{3mm}
\caption{\label{fig:WWqqlvmin13} Invariant $WW$ mass distributions for the signal 
process $gg\to H\to W^-W^+\to \ell\bar{\nu}_\ell q_u \bar{q}_d$ 
and including its interference with the background 
in $pp$ collisions at $\sqrt{s}=13$ TeV.
Minimal cuts are applied (see main text).  
Other details as in figure \ref{fig:WWqqlvmin}.}
\end{figure}

\begin{figure}[tb]
\centering
\includegraphics[width = 0.9\textwidth]{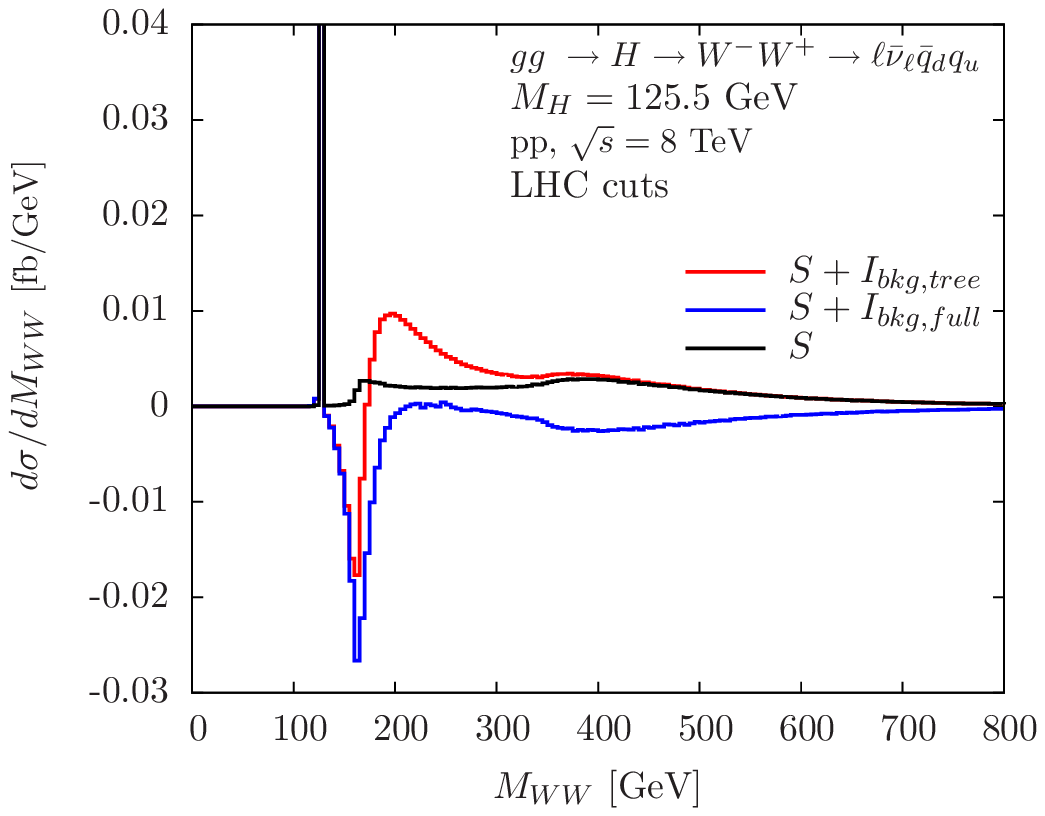}\hfill\hspace{3mm}
\caption{\label{fig:WWqqlvLHC}Invariant $WW$ mass distributions for the signal 
process $gg\to H\to W^-W^+\to \ell\bar{\nu}_\ell q_u \bar{q}_d$ 
and including its interference with the background in $pp$ collisions at $\sqrt{s}=8$ TeV.
LHC cuts are applied (see main text). Other details as in figure 
\ref{fig:WWqqlvmin}.}
\end{figure}

\begin{figure}[tb]
\centering
\includegraphics[width = 0.9\textwidth]{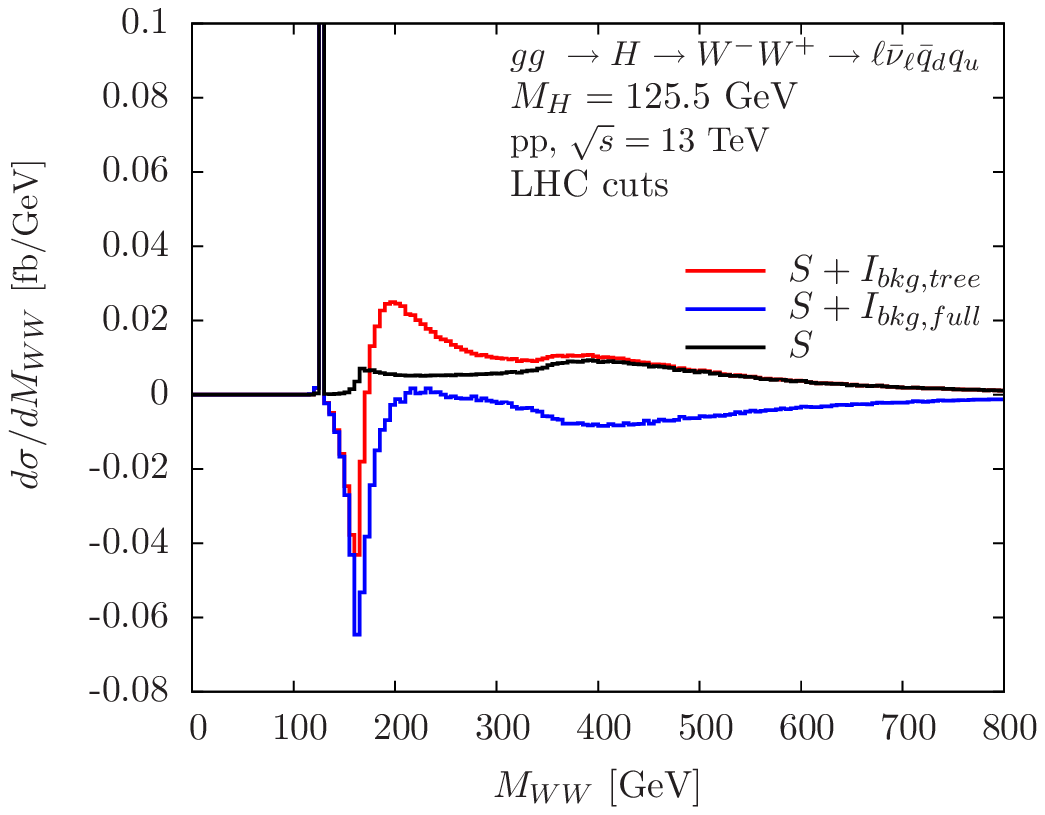}\hfill\hspace{3mm}
\caption{\label{fig:WWqqlvLHC13}Invariant $WW$ mass distributions for the signal 
process $gg\to H\to W^-W^+\to \ell\bar{\nu}_\ell q_u \bar{q}_d$ 
and including its interference with the background in $pp$ collisions at $\sqrt{s}=13$ TeV.
LHC cuts are applied (see main text). Other details as in figure 
\ref{fig:WWqqlvmin}.}
\end{figure}

\begin{figure}[tb]
\centering
\includegraphics[width = 0.9\textwidth]{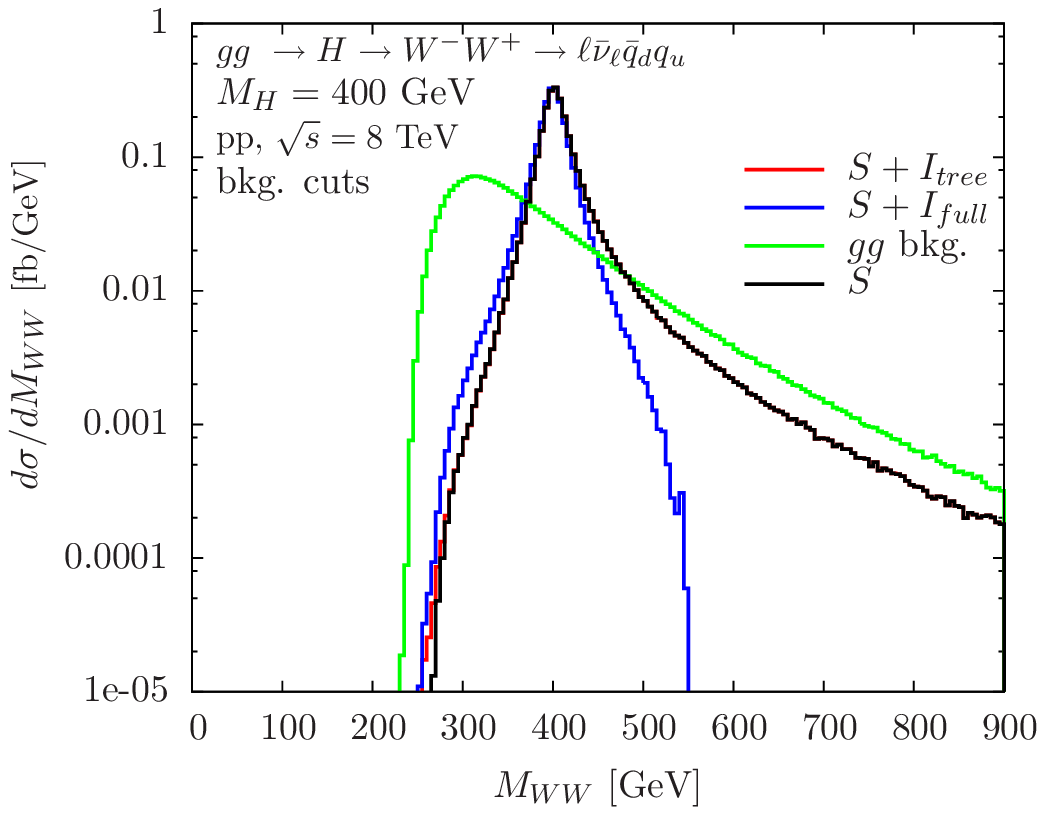}\hfill\hspace{3mm}
\caption{\label{fig:WWqqlvbkg_bg}
Invariant $WW$ mass distributions for the signal process $gg\to H\to W^-W^+\to \ell\bar{\nu}_\ell q_u \bar{q}_d$ 
and including its interference with the background in $pp$ collisions at $\sqrt{s}=8$ TeV
with background suppression cuts for a $400$ GeV SM Higgs boson (see main text).
The $gg$ background is also displayed. Other details as in figure 
\ref{fig:WWqqlvmin}.}
\end{figure}

\begin{figure}[tb]
\centering
\includegraphics[width = 0.9\textwidth]{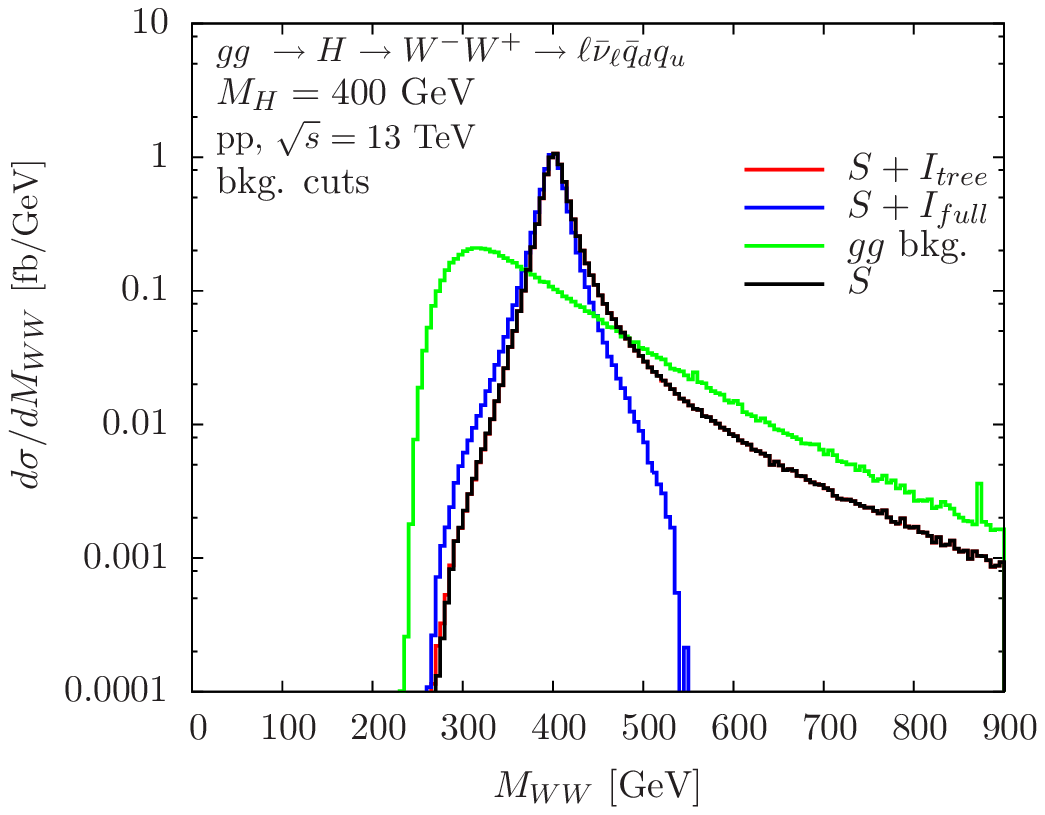}\hfill\hspace{3mm}
\caption{\label{fig:WWqqlvbkg_bg13}
Invariant $WW$ mass distributions for the signal process $gg\to H\to W^-W^+\to \ell\bar{\nu}_\ell q_u \bar{q}_d$ 
and including its interference with the background in $pp$ collisions at $\sqrt{s}=13$ TeV
with background suppression cuts for a $400$ GeV SM Higgs boson (see main text).
Other details as in figure 
\ref{fig:WWqqlvbkg_bg}.}
\end{figure}

\begin{figure}[tb]
\centering
\includegraphics[width = 0.9\textwidth]{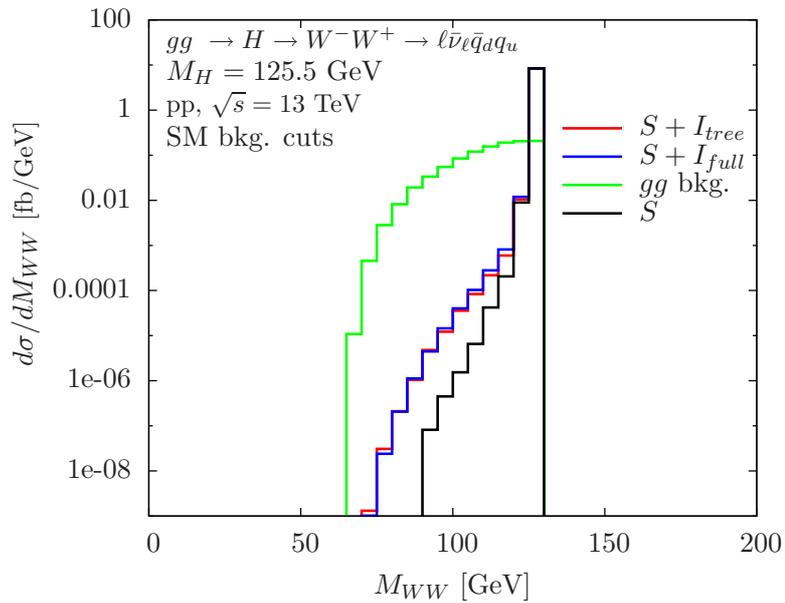}\hfill\hspace{3mm}
\caption{\label{fig:WWlvqqsmbkg} Invariant $WW$ mass distributions for the signal process $gg\to H\to W^-W^+\to \ell\bar{\nu}_\ell q_u \bar{q}_d$ 
and including its interference with the background in $pp$ collisions at $\sqrt{s}=13$ TeV with background 
suppression cuts for a $125.5$ GeV Higgs boson at $\sqrt{s}\approx 14$ TeV 
(see main text). The $gg$ background is also displayed.  Other details as in figure 
\ref{fig:WWqqlvmin}.}
\end{figure}

\begin{figure}[tb]
\centering
\includegraphics[width = 0.9\textwidth]{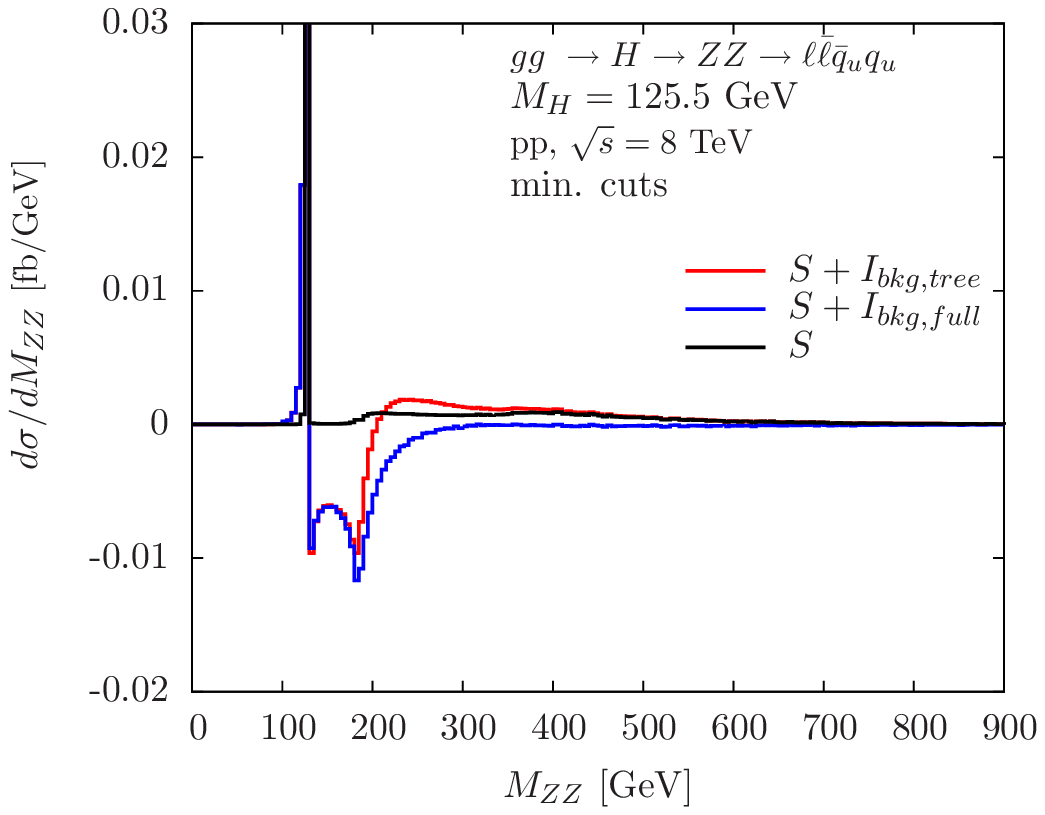}\hfill\hspace{3mm}
\caption{\label{fig:ZZlluumin} Invariant $ZZ$ mass distributions for the signal process $gg\to H\to ZZ\to \ell\bar{\ell} q_u \bar{q}_u$
and including its interference with the background in $pp$ collisions at $\sqrt{s}=8$ TeV.
Minimal cuts are applied (see main text). $\gamma^\ast$ background contributions are included. 
Other details as in figure \ref{fig:WWqqlvmin}.}
\end{figure}

\begin{figure}[tb]
\centering
\includegraphics[width = 0.9\textwidth]{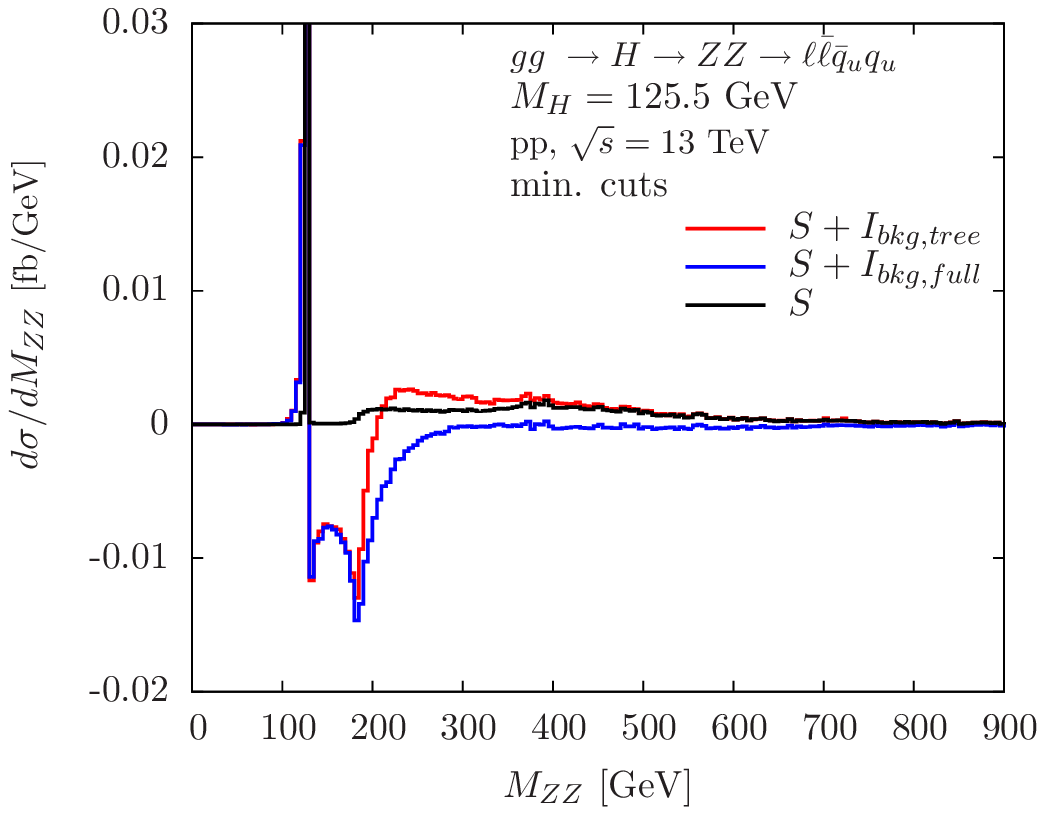}\hfill\hspace{3mm}
\caption{\label{fig:ZZlluumin13} Invariant $ZZ$ mass distributions for the signal process $gg\to H\to ZZ\to \ell\bar{\ell} q_u \bar{q}_u$
and including its interference with the background in $pp$ collisions at $\sqrt{s}=13$ TeV.
Minimal cuts are applied (see main text). Other details as in figure \ref{fig:ZZlluumin}.}
\end{figure}

\begin{figure}[tb]
\centering
\includegraphics[width = 0.9\textwidth]{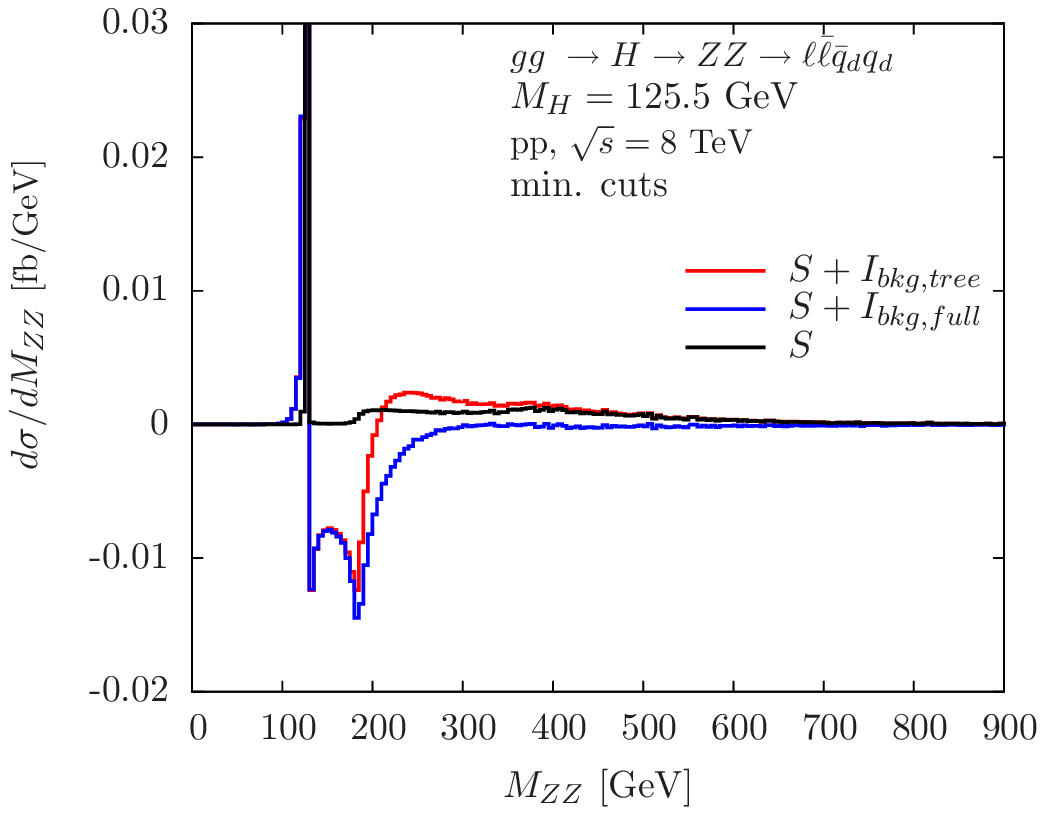}\hfill\hspace{3mm}
\caption{\label{fig:ZZllddmin} Invariant $ZZ$ mass distributions for the signal 
process 
$gg\to H\to ZZ\to \ell\bar{\ell} q_d \bar{q}_d$
and including its interference with the background in $pp$ collisions at $\sqrt{s}=8$ TeV.
Minimal cuts are applied (see main text). Other details as in figure 
\ref{fig:ZZlluumin}.}
\end{figure}

\begin{figure}[tb]
\centering
\includegraphics[width = 0.9\textwidth]{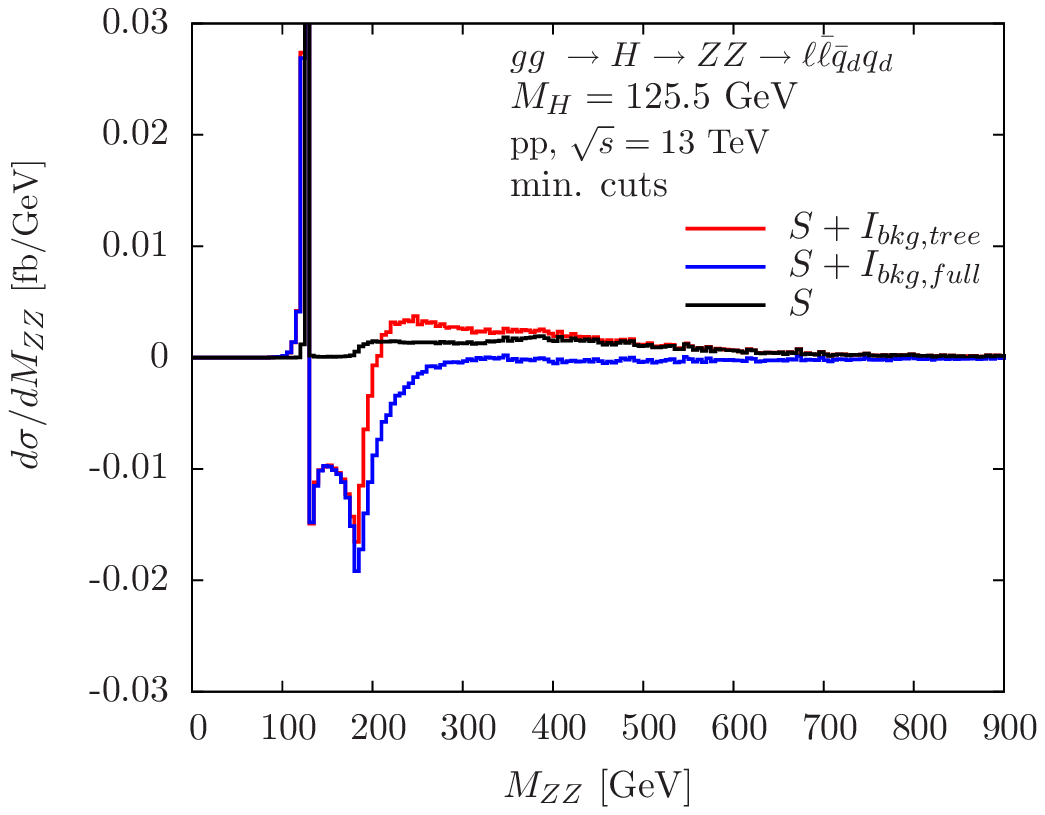}\hfill\hspace{3mm}
\caption{\label{fig:ZZllddmin13} Invariant $ZZ$ mass distributions for the signal 
process 
$gg\to H\to ZZ\to \ell\bar{\ell} q_d \bar{q}_d$
and including its interference with the background in $pp$ collisions at $\sqrt{s}=13$ TeV.
Minimal cuts are applied (see main text). Other details as in figure 
\ref{fig:ZZlluumin}.}
\end{figure}

\begin{figure}
\centering
\includegraphics[width = 0.9\textwidth]{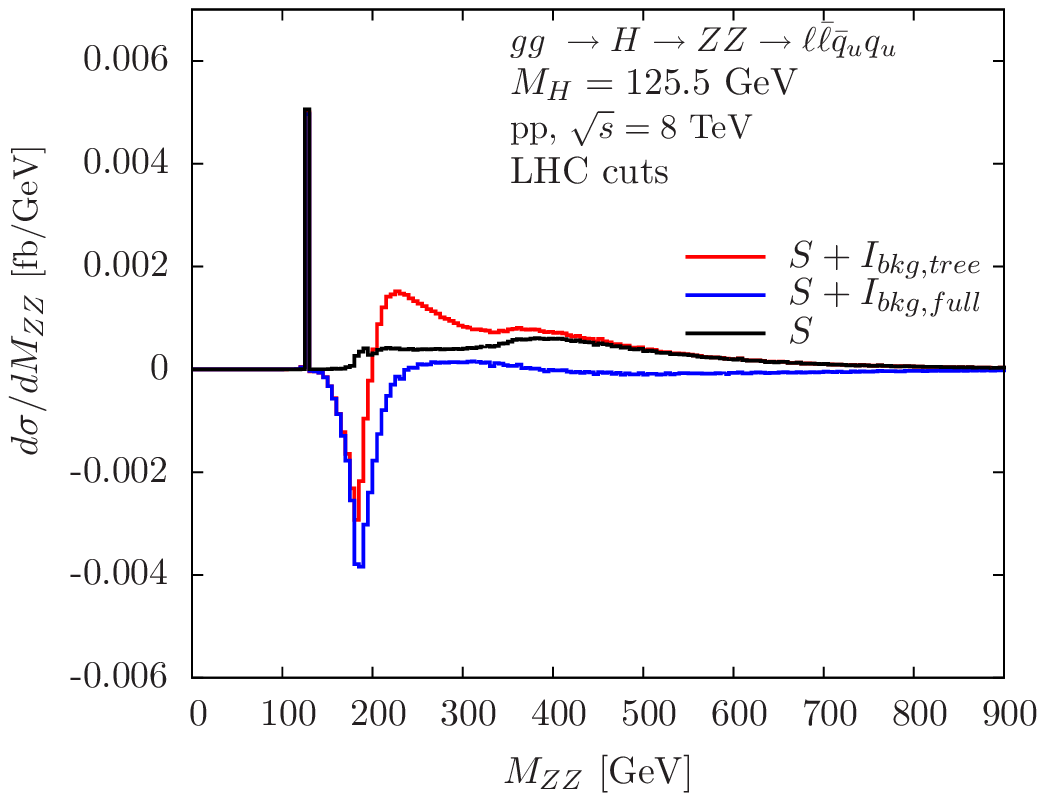}\hfill\hspace{3mm}
\caption{\label{fig:ZZlluuLHC} Invariant $ZZ$ mass distributions for the signal process $gg\to H\to ZZ\to \ell\bar{\ell} q_u \bar{q}_u$
and including its interference with the background in $pp$ collisions at $\sqrt{s}=8$ TeV.
LHC cuts are applied (see main text). Other details as in figure 
\ref{fig:ZZlluumin}.}
\end{figure}

\begin{figure}
\centering
\includegraphics[width = 0.9\textwidth]{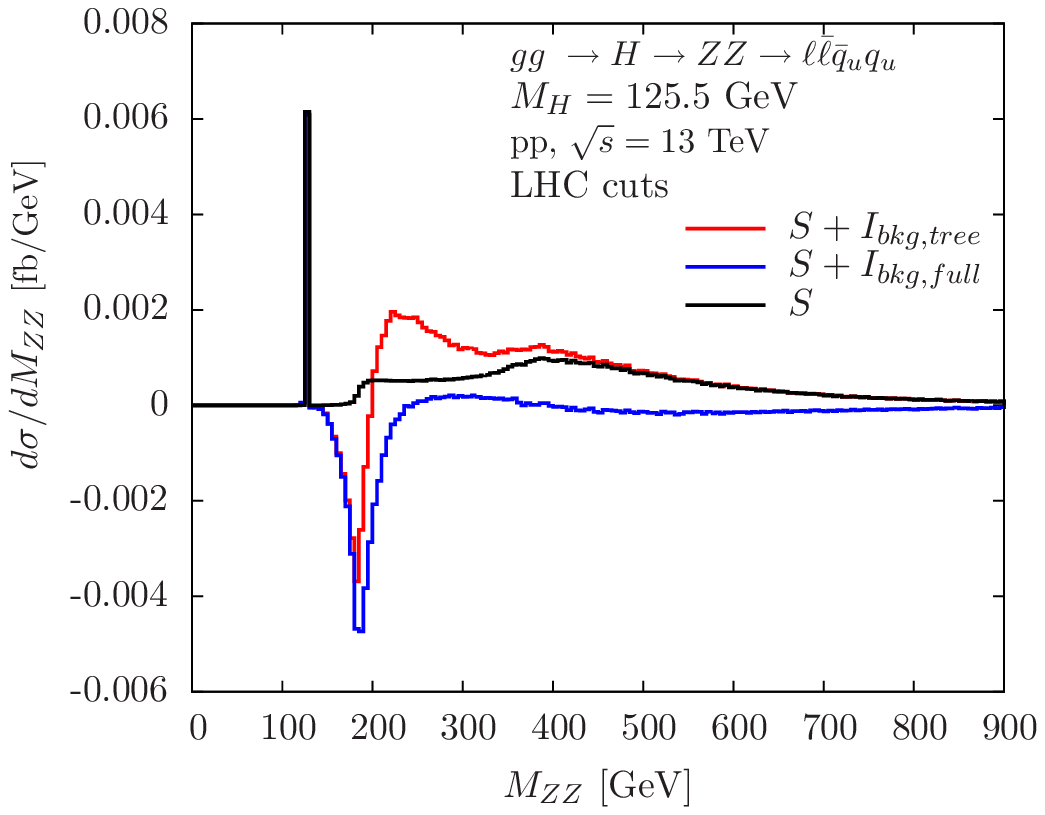}\hfill\hspace{3mm}
\caption{\label{fig:ZZlluuLHC13} Invariant $ZZ$ mass distributions for the signal process $gg\to H\to ZZ\to \ell\bar{\ell} q_u \bar{q}_u$
and including its interference with the background in $pp$ collisions at $\sqrt{s}=13$ TeV.
LHC cuts are applied (see main text). Other details as in figure 
\ref{fig:ZZlluumin}.}
\end{figure}

\begin{figure}[tb]
\centering
\includegraphics[width = 0.9\textwidth]{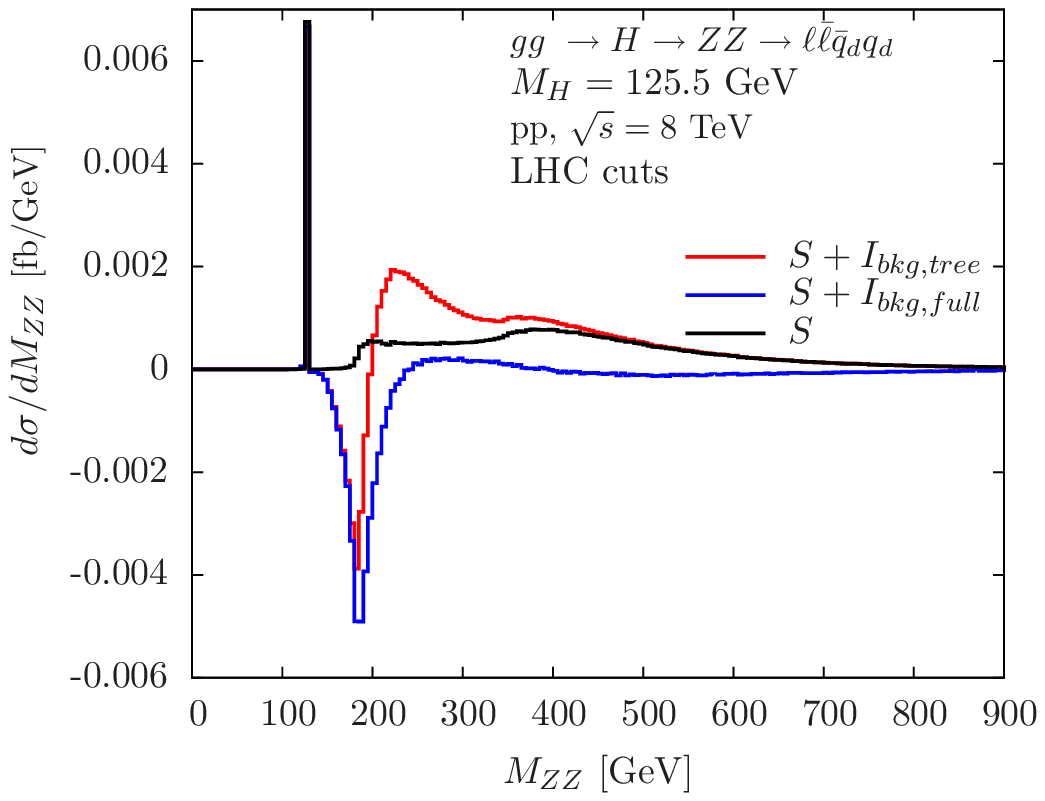}\hfill\hspace{3mm}
\caption{\label{fig:ZZllddLHC} Invariant $ZZ$ mass distributions for the signal 
process 
$gg\to H\to ZZ\to \ell\bar{\ell} q_d \bar{q}_d$
and including its interference with the background in $pp$ collisions at $\sqrt{s}=8$ TeV.
LHC cuts are applied (see main text). Other details as in figure 
\ref{fig:ZZlluumin}.}
\end{figure}

\begin{figure}[tb]
\centering
\includegraphics[width = 0.9\textwidth]{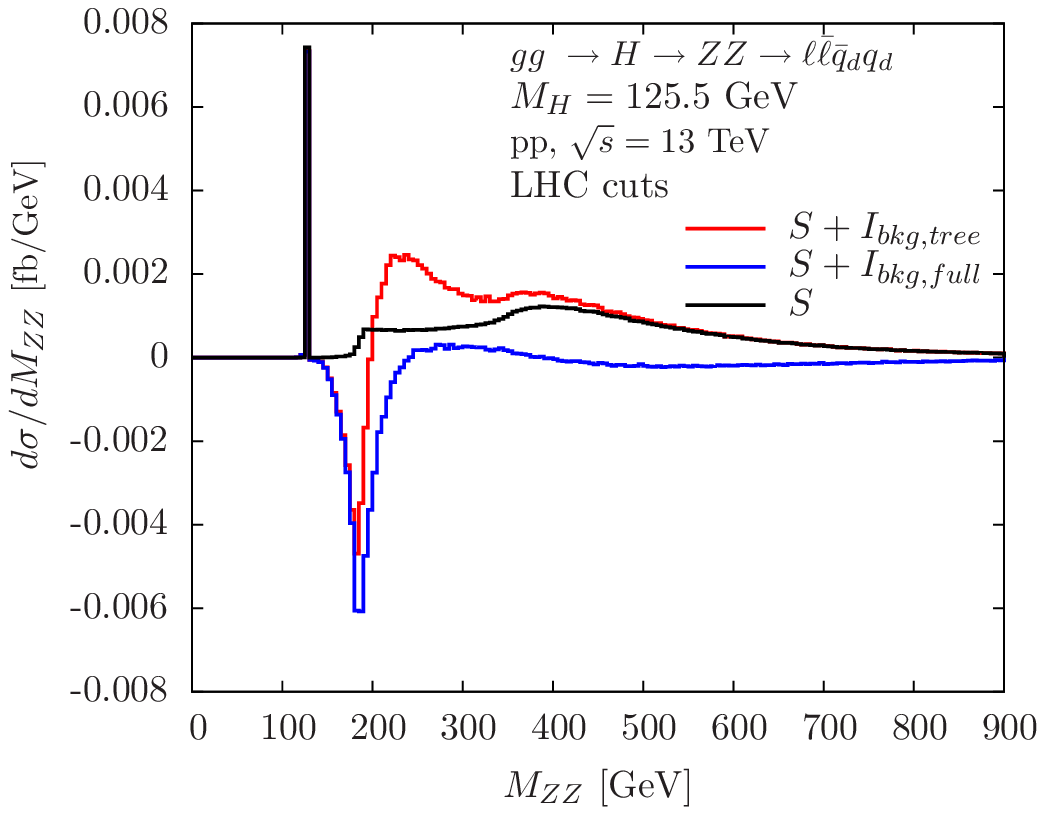}\hfill\hspace{3mm}
\caption{\label{fig:ZZllddLHC13} Invariant $ZZ$ mass distributions for the signal 
process 
$gg\to H\to ZZ\to \ell\bar{\ell} q_d \bar{q}_d$
and including its interference with the background in $pp$ collisions at $\sqrt{s}=13$ TeV.
LHC cuts are applied (see main text). Other details as in figure 
\ref{fig:ZZlluumin}.}
\end{figure}

\begin{figure}[tb]
\centering
\includegraphics[width = 0.9\textwidth]{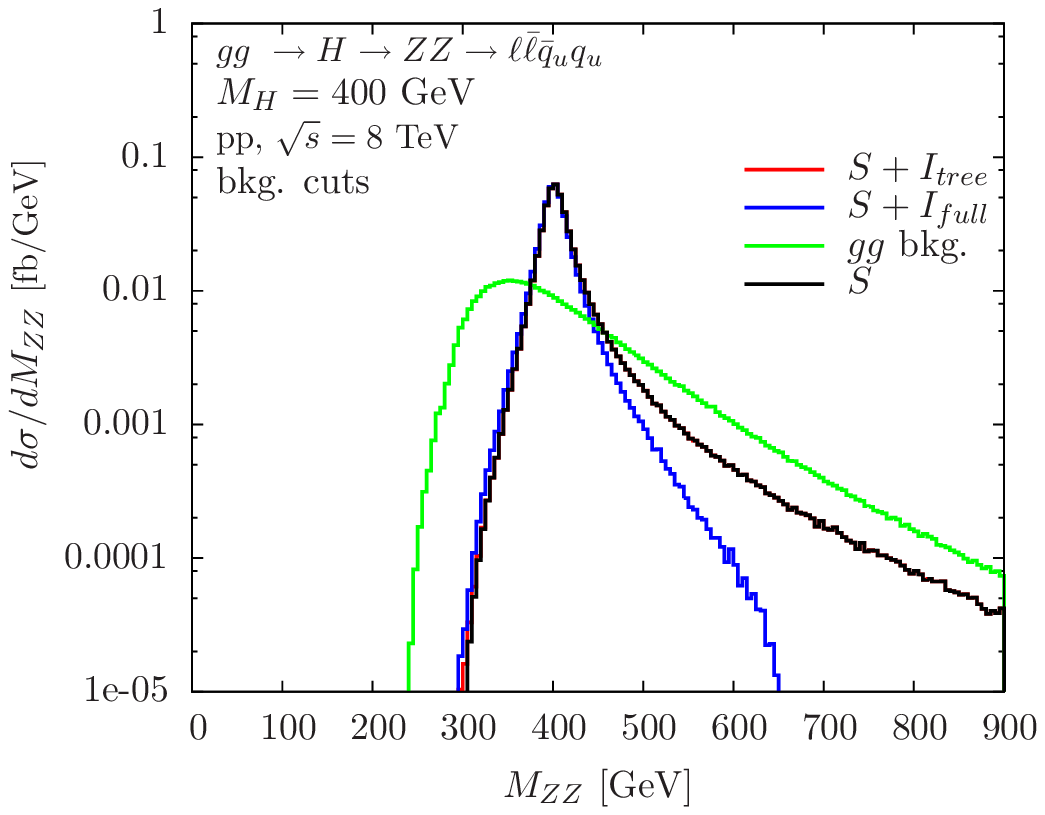}\hfill\hspace{3mm}
\caption{\label{fig:ZZlluubkg}
Invariant $ZZ$ mass distributions for the signal process 
$gg\to H\to ZZ\to \ell\bar{\ell} q_u \bar{q}_u$
and including its interference with the background in $pp$ collisions at $\sqrt{s}=8$ TeV
with background suppression cuts for a $400$ GeV SM Higgs boson (see main text).
The $gg$ background is also displayed. Other details as in figure 
\ref{fig:ZZlluumin}.}
\end{figure}

\begin{figure}[tb]
\centering
\includegraphics[width = 0.9\textwidth]{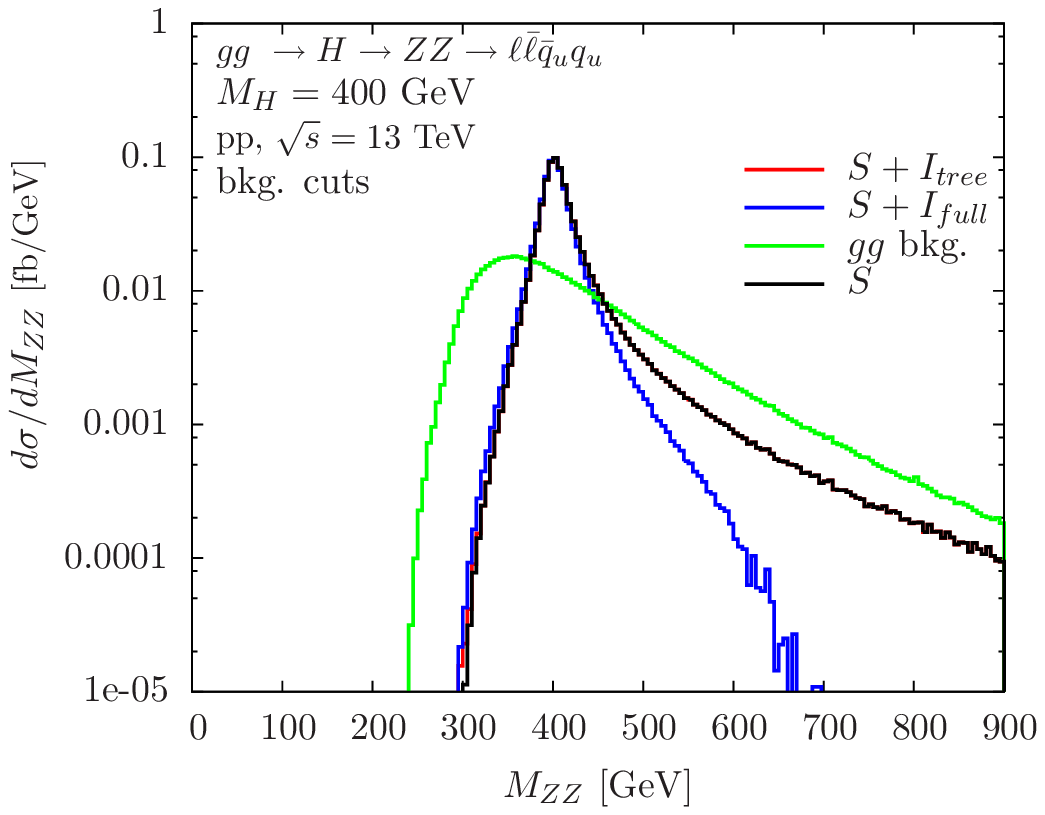}\hfill\hspace{3mm}
\caption{\label{fig:ZZlluubkg13}
Invariant $ZZ$ mass distributions for the signal process 
$gg\to H\to ZZ\to \ell\bar{\ell} q_u \bar{q}_u$
and including its interference with the background in $pp$ collisions at $\sqrt{s}=13$ TeV
with background suppression cuts for a $400$ GeV SM Higgs boson (see main text).
The $gg$ background is also displayed. Other details as in figure 
\ref{fig:ZZlluumin}.}
\end{figure}

\begin{figure}[tb]
\centering
\includegraphics[width = 0.9\textwidth]{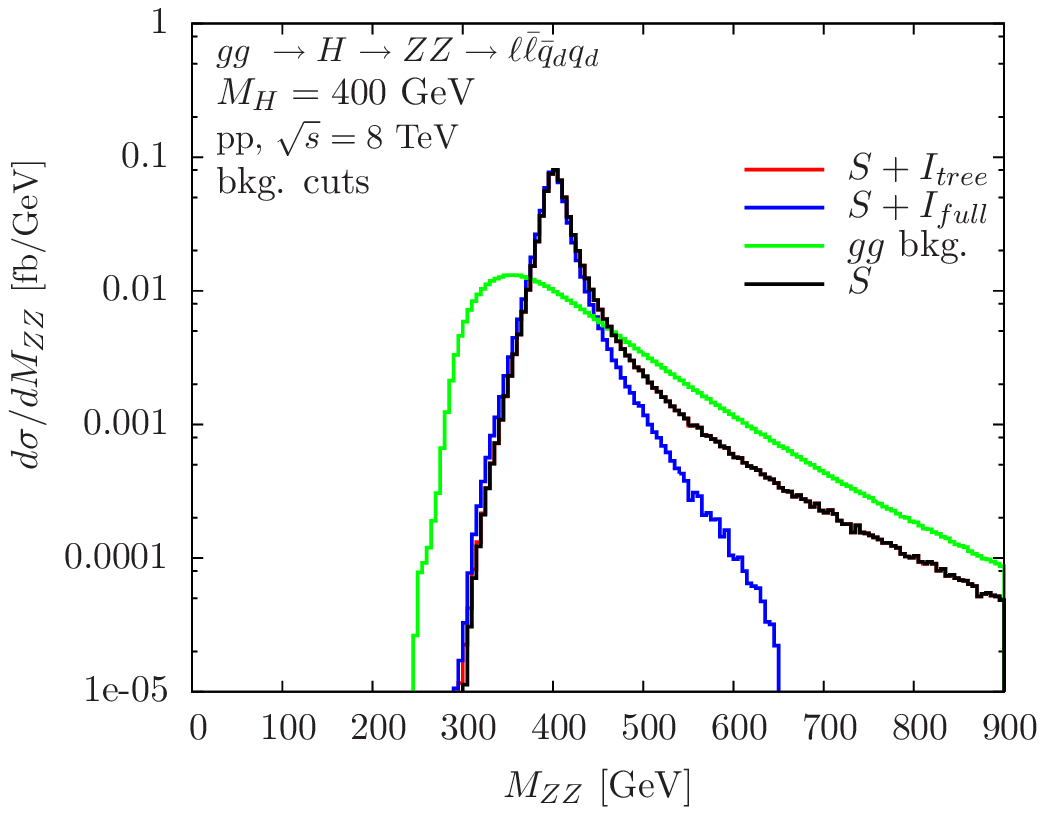}\hfill\hspace{3mm}
\caption{\label{fig:ZZllddbkg}
Invariant $ZZ$ mass distributions for the signal process 
$gg\to H\to ZZ\to \ell\bar{\ell} q_d \bar{q}_d$ 
and including its interference with the background in $pp$ collisions at $\sqrt{s}=8$ TeV
with background suppression cuts for a $400$ GeV SM Higgs boson (see main text).
The $gg$ background is also displayed. Other details as in figure 
\ref{fig:ZZlluumin}.}
\end{figure}

\begin{figure}[tb]
\centering
\includegraphics[width = 0.9\textwidth]{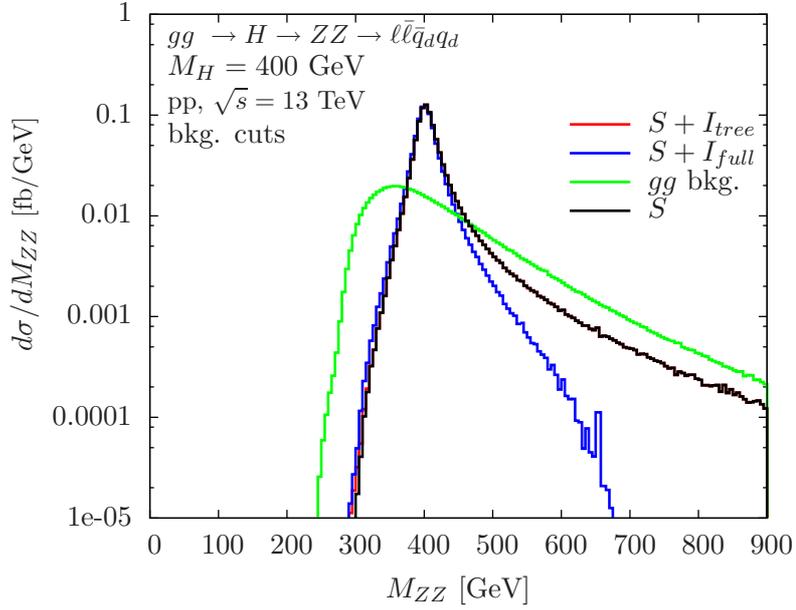}\hfill\hspace{3mm}
\caption{\label{fig:ZZllddbkg13}
Invariant $ZZ$ mass distributions for the signal process 
$gg\to H\to ZZ\to \ell\bar{\ell} q_d \bar{q}_d$ 
and including its interference with the background in $pp$ collisions at $\sqrt{s}=13$ TeV
with background suppression cuts for a $400$ GeV SM Higgs boson (see main text).
The $gg$ background is also displayed. Other details as in figure 
\ref{fig:ZZlluumin}.}
\end{figure}

In figures \ref{fig:WWqqlvmin}--\ref{fig:ZZllddbkg13}, the corresponding differential distributions for the invariant mass of the $\ell\bar{\nu}q\bar{q}$ ($V=W$) and $\ell\bar{\ell}q\bar{q}$ ($V=Z$) systems, denoted by $M_{VV}$ in the plots, are 
shown. In addition to the observations made above at the integrated cross section level, interesting information can also be extracted by studying the $M_{VV}$ distributions.  Firstly, we note that all figures are dominated by a sharp resonance peak at the Higgs mass induced by the signal.  On the resonance peak, interference effects are 
negligible \cite{Kauer:2012hd}.  Moving away from the resonance peak increases the importance of the interference effects which rapidly overtake the signal in size.  

A general observation for the light Higgs case is the appearance of various thresholds in the $M_{VV}$ distribution. The $2M_V$ threshold arises in both the signal and loop background amplitudes when the two weak bosons are produced on-shell. A second threshold occurs at $2M_t$, again relevant for the signal and loop background amplitudes: when the top quarks in the loop are produced on-shell the amplitude acquires an imaginary part. These loop amplitude thresholds are well known and have been extensively discussed in the literature.

Another feature that we observe in all cases by studying the interference contributions individually is that the tree-level background displays rather sharp dips and peaks. The sign of the tree-level interference changes, often leading to sizeable cancellations between regions of different invariant mass and consequently to a reduced contribution to the integrated cross section. A similar cancellation occurs for the loop-induced background for the heavy Higgs case, as the interference changes sign at $M_H$.
In this case the total interference contribution to the signal cross section 
is reduced to 7--10\%.


\clearpage


\clearpage


\section{Conclusions\label{sec:conclusions}}

The semileptonic $H\to WW$ and $H\to ZZ$ channels provide important additional information in SM light Higgs measurements and heavy Higgs searches. In this work, we have studied the signal-background interference for $H\to WW \to \ell \nu  q \bar{q}' $ and $H\to ZZ \to \ell\bar{\ell} q\bar{q}$ in gluon fusion at the LHC, computing both the tree-level and quark-loop $gg$ background contributions to the interference. The former is enhanced by $1/e^2$ relative to the latter. The results have been obtained using two independent implementations in the public codes \textsf{gg2VV} and {\sc MadGraph5\_aMC@NLO}, which show excellent agreement.  We have considered light and heavy Higgs masses with minimal and realistic experimental selection cuts.

We find that the interference of the Higgs signal with the tree-level background is 
suppressed compared to its interference with the loop-induced $VV$ 
background.  This holds for light and heavy Higgs masses as well as minimal, LHC 
and Higgs search cuts.  Generalising this result, we conclude that 
higher-order background contributions can induce leading interference effects.
Consequently, precision calculations of interference effects are well motivated.

The impact of the full signal-background interference on integrated signal cross sections can range from $\mathcal{O}(0.1\%)$ to $\mathcal{O}(1)$ depending on the applied selection cuts.  We show that for the light SM Higgs with appropriate background suppression cuts, the total effect of the signal-background interference on the integrated cross section can be reduced to the per mille level in the $WW$ channel. Similarly, background suppression cuts for a heavy Higgs boson can lower the interference to 
10\% or less for the $WW$ and $ZZ$ channel.

In addition to the integrated cross section study, we have calculated differential results and discussed important features in the invariant mass distributions, particularly in $M_{VV}$ regions away from the (light or heavy) Higgs mass. In these off-shell regions interference effects dominate over the signal and therefore should be taken into account for more exclusive selection cuts or analysis methods.  The partial cancellation of positive and negative interference contributions, which mitigates interference effects for sufficiently inclusive selections, in general does not apply to more exclusive selections or differential distributions.


\acknowledgments
N.K.\ would like to thank the Galileo Galilei Institute for Theoretical Physics
for hospitality and the INFN for partial support during the preparation of
this paper. C.O.\ would like to thank the Department of Physics, Royal Holloway, 
University of London for supplementary financial support. E.V.\ would like to thank V. Hirschi and O. Mattelaer for valuable assistance, and F. Maltoni for useful discussions. This work was supported by STFC grants ST/J000485/1, ST/J005010/1 
and ST/L000512/1 and in part by the European Union as part of the FP7 Marie Curie Initial Training Network MCnetITN  (PITN-GA-2012-315877). 




\begin{thebibliography}{22}


\bibitem{Aad:2012tfa} 
  G.~Aad {\it et al.}  [ATLAS Collaboration],
  \emph{Observation of a new particle in the search for the Standard Model Higgs boson with the ATLAS detector at the LHC},
  Phys.\ Lett.\ B {\bf 716} (2012) 1
  [arXiv:1207.7214 [hep-ex]].

\bibitem{Chatrchyan:2012ufa} 
  S.~Chatrchyan {\it et al.}  [CMS Collaboration],
  \emph{Observation of a new boson at a mass of 125 GeV with the CMS experiment at the LHC},
  Phys.\ Lett.\ B {\bf 716} (2012) 30
  [arXiv:1207.7235 [hep-ex]].

\bibitem{Higgs:1964ia}
  P.~W.~Higgs,
  \emph{Broken symmetries, massless particles and gauge fields},
  Phys.\ Lett.\  {\bf 12} (1964) 132.

\bibitem{Higgs:1964pj}
  P.~W.~Higgs,
  \emph{Broken symmetries and the masses of gauge bosons},
  Phys.\ Rev.\ Lett.\  {\bf 13} (1964) 508.


\bibitem{Englert:1964et}
  F.~Englert and R.~Brout,
  \emph{Broken symmetry and the mass of gauge vector mesons},
  Phys.\ Rev.\ Lett.\  {\bf 13} (1964) 321.

\bibitem{Guralnik:1964eu}
  G.~S.~Guralnik, C.~R.~Hagen and T.~W.~B.~Kibble,
  \emph{Global conservation laws and massless particles},
  Phys.\ Rev.\ Lett.\  {\bf 13} (1964) 585.



  
\bibitem{Georgi:1977gs}
  H.~M.~Georgi, S.~L.~Glashow, M.~E.~Machacek and D.~V.~Nanopoulos,
  \emph{Higgs bosons from two-gluon annihilation in proton-proton collisions},
  Phys.\ Rev.\ Lett.\  {\bf 40} (1978) 692.

\bibitem{Dawson:1990zj}
  S.~Dawson,
  \emph{Radiative corrections to Higgs boson production},
  Nucl.\ Phys.\ B {\bf 359} (1991) 283.

\bibitem{Djouadi:1991tka}
  A.~Djouadi, M.~Spira and P.~M.~Zerwas,
  \emph{Production of Higgs bosons in proton colliders: QCD corrections},
  Phys.\ Lett.\ B {\bf 264} (1991) 440.

\bibitem{Graudenz:1992pv}
  D.~Graudenz, M.~Spira and P.~M.~Zerwas,
  \emph{QCD corrections to Higgs boson production at proton-proton colliders},
  Phys.\ Rev.\ Lett.\  {\bf 70} (1993) 1372.

\bibitem{Spira:1995rr}
  M.~Spira, A.~Djouadi, D.~Graudenz and P.~M.~Zerwas,
  \emph{Higgs boson production at the LHC},
  Nucl.\ Phys.\ B {\bf 453} (1995) 17
  [hep-ph/9504378].

\bibitem{Harlander:2002wh}
  R.~V.~Harlander and W.~B.~Kilgore,
  \emph{Next-to-next-to-leading order Higgs production at hadron colliders},
  Phys.\ Rev.\ Lett.\  {\bf 88} (2002) 201801
  [hep-ph/0201206].

\bibitem{Anastasiou:2002yz}
  C.~Anastasiou and K.~Melnikov,
  \emph{Higgs boson production at hadron colliders in NNLO QCD},
  Nucl.\ Phys.\  B {\bf 646} (2002) 220
  [arXiv:hep-ph/0207004].

\bibitem{Ravindran:2003um}
  V.~Ravindran, J.~Smith and W.~L.~van Neerven,
  \emph{NNLO corrections to the total cross section for Higgs boson production in hadron-hadron collisions},
  Nucl.\ Phys.\  B {\bf 665} (2003) 325
  [arXiv:hep-ph/0302135].
      
\bibitem{Anastasiou:2015ema}
  C.~Anastasiou, C.~Duhr, F.~Dulat, F.~Herzog and B.~Mistlberger,
  \emph{Higgs boson gluon-fusion production in QCD at three loops},
  Phys.\ Rev.\ Lett.\  {\bf 114} (2015) 212001
  [arXiv:1503.06056 [hep-ph]].


\bibitem{Dittmaier:2011ti}
  S.~Dittmaier {\it et al.},
  \emph{Handbook of LHC Higgs cross sections: 1.\ Inclusive observables},
  arXiv:1101.0593 [hep-ph].

\bibitem{Dittmaier:2012vm}
  S.~Dittmaier {\it et al.},
  \emph{Handbook of LHC Higgs cross sections: 2.\ Differential distributions},
  arXiv:1201.3084 [hep-ph].

\bibitem{Heinemeyer:2013tqa}
  S.~Heinemeyer {\it et al.}, 
  \emph{Handbook of LHC Higgs cross sections: 3.\ Higgs properties},
  arXiv:1307.1347 [hep-ph].


\bibitem{Glover:1988fe}
  E.~W.~N.~Glover and J.~J.~van der Bij,
  \emph{Vector boson pair production via gluon fusion},
  Phys.\ Lett.\ B {\bf 219} (1989) 488.

\bibitem{Glover:1988rg}
  E.~W.~N.~Glover and J.~J.~van der Bij,
  \emph{$Z$-boson pair production via gluon fusion},
  Nucl.\ Phys.\ B {\bf 321} (1989) 561.

\bibitem{Binoth:2006mf}
  T.~Binoth, M.~Ciccolini, N.~Kauer and M.~Kramer,
  \emph{Gluon-induced W-boson pair production at the LHC},
  JHEP {\bf 0612} (2006) 046
  [hep-ph/0611170].

\bibitem{Campbell:2011cu}
  J.~M.~Campbell, R.~K.~Ellis and C.~Williams,
  \emph{Gluon-gluon contributions to $W^+W^-$ production and Higgs interference effects},
  JHEP {\bf 1110} (2011)  005
  [arXiv:1107.5569 [hep-ph]].

\bibitem{Kauer:2012ma}
  N.~Kauer,
  \emph{Signal-background interference in $gg \to H \to VV$},
  PoS RADCOR {\bf 2011} (2011) 027
  [arXiv:1201.1667 [hep-ph]].

\bibitem{Passarino:2012ri}
  G.~Passarino,
  \emph{Higgs interference effects in $gg \to ZZ$ and their uncertainty},
  JHEP {\bf 1208} (2012) 146
  [arXiv:1206.3824 [hep-ph]].

\bibitem{Kauer:2012hd}
  N.~Kauer and G.~Passarino,
  \emph{Inadequacy of zero-width approximation for a light Higgs boson signal},
  JHEP {\bf 1208} (2012) 116
  [arXiv:1206.4803 [hep-ph]].

\bibitem{Bonvini:2013jha}
  M.~Bonvini, F.~Caola, S.~Forte, K.~Melnikov and G.~Ridolfi,
  \emph{Signal-background interference effects for $gg \to H \to W^+ W^-$ beyond leading order},
  Phys.\ Rev.\ D {\bf 88} (2013) 3,  034032
  [arXiv:1304.3053 [hep-ph]].

\bibitem{Kauer:2013qba}
  N.~Kauer,
  \emph{Interference effects for $H \to WW/ZZ \to \ell\bar{\nu}_\ell\bar{\ell}\nu_\ell$ searches in gluon fusion at the LHC},
  JHEP {\bf 1312} (2013) 082
  [arXiv:1310.7011 [hep-ph]].

\bibitem{Campbell:2013una}
  J.~M.~Campbell, R.~K.~Ellis and C.~Williams,
  \emph{Bounding the Higgs width at the LHC using full analytic results for $gg \to e^- e^+ \mu^- \mu^+$},
  JHEP {\bf 1404} (2014) 060
  [arXiv:1311.3589 [hep-ph]].

\bibitem{Moult:2014pja}
  I.~Moult and I.~W.~Stewart,
  \emph{Jet vetoes interfering with $H \to WW$},
  JHEP {\bf 1409} (2014) 129
  [arXiv:1405.5534 [hep-ph]].

\bibitem{Ellis:2014yca}
  J.~M.~Campbell, R.~K.~Ellis and C.~Williams,
  \emph{Bounding the Higgs width at the LHC},
  PoS LL {\bf 2014} (2014) 008
  [arXiv:1408.1723 [hep-ph]].

\bibitem{Campanario:2012bh}
  F.~Campanario, Q.~Li, M.~Rauch and M.~Spira,
  \emph{ZZ+jet production via gluon fusion at the LHC},
  JHEP {\bf 1306} (2013) 069
  [arXiv:1211.5429 [hep-ph]].

\bibitem{Campbell:2014gua}
  J.~M.~Campbell, R.~K.~Ellis, E.~Furlan and R.~Rontsch,
  \emph{Interference effects for Higgs boson mediated $Z$-pair plus jet production},
  Phys.\ Rev.\ D {\bf 90} (2014) 9,  093008
  [arXiv:1409.1897 [hep-ph]].

\bibitem{Li:2015jva}
  C.~S.~Li, H.~T.~Li, D.~Y.~Shao and J.~Wang,
  \emph{Soft gluon resummation in the signal-background interference process of $gg\ (\to h^*) \to ZZ$},
  JHEP {\bf 1508} (2015) 065
  [arXiv:1504.02388 [hep-ph]].


\bibitem{Cascioli:2013gfa}
  F.~Cascioli, S.~Hoche, F.~Krauss, P.~Maierhofer, S.~Pozzorini and F.~Siegert,
  \emph{Precise Higgs-background predictions: merging NLO QCD and squared quark-loop corrections to four-lepton + 0,1 jet production},
  JHEP {\bf 1401} (2014) 046
  [arXiv:1309.0500 [hep-ph]].
  
\bibitem{Liebler:2015aka}
  S.~Liebler, G.~Moortgat-Pick and G.~Weiglein,
  \emph{Off-shell effects in Higgs processes at a linear collider and implications for the LHC},
  JHEP {\bf 1506} (2015) 093
  [arXiv:1502.07970 [hep-ph]].


\bibitem{Caola:2015ila}
  F.~Caola, J.~M.~Henn, K.~Melnikov, A.~V.~Smirnov and V.~A.~Smirnov,
  \emph{Two-loop helicity amplitudes for the production of two off-shell electroweak bosons in gluon fusion},
  JHEP {\bf 1506} (2015) 129
  [arXiv:1503.08759 [hep-ph]].

\bibitem{vonManteuffel:2015msa}
  A.~von Manteuffel and L.~Tancredi,
  \emph{The two-loop helicity amplitudes for $gg \to V_1 V_2 \to 4~\mathrm{leptons}$},
  JHEP {\bf 1506} (2015) 197
  [arXiv:1503.08835 [hep-ph]].

\bibitem{Melnikov:2015laa}
  K.~Melnikov and M.~Dowling,
  \emph{Production of two Z-bosons in gluon fusion in the heavy top quark approximation},
  Phys.\ Lett.\ B {\bf 744} (2015) 43
  [arXiv:1503.01274 [hep-ph]].


\bibitem{Cascioli:2014yka}
  F.~Cascioli, T.~Gehrmann, M.~Grazzini, S.~Kallweit, P.~Maierhofer, A.~von Manteuffel, S.~Pozzorini, D.~Rathlev {\it et al.},
  \emph{ZZ production at hadron colliders in NNLO QCD},
  Phys.\ Lett.\ B {\bf 735} (2014) 311
  [arXiv:1405.2219 [hep-ph]].

\bibitem{Gehrmann:2014fva}
  T.~Gehrmann, M.~Grazzini, S.~Kallweit, P.~Maierhofer, A.~von Manteuffel, S.~Pozzorini, D.~Rathlev and L.~Tancredi,
  \emph{$W^+W^-$ production at hadron colliders in next-to-next-to-leading order QCD},
  Phys.\ Rev.\ Lett.\  {\bf 113} (2014) 21,  212001
  [arXiv:1408.5243 [hep-ph]].


\bibitem{Aad:2012oxa}
  G.~Aad {\it et al.}  [ATLAS Collaboration],
  \emph{Search for a Standard Model Higgs boson in the mass range 200--600 GeV in the $H \to ZZ \to \ell^+ \ell^- q \bar{q}$ decay channel with the ATLAS detector},
  Phys.\ Lett.\ B {\bf 717} (2012) 70
  [arXiv:1206.2443 [hep-ex]].

\bibitem{Aad:2012me}
  G.~Aad {\it et al.}  [ATLAS Collaboration],
  \emph{Search for the Higgs boson in the $H \to W W \to l \nu jj$ decay channel at $\sqrt{s}=7$ TeV with the ATLAS detector},
  Phys.\ Lett.\ B {\bf 718} (2012) 391
  [arXiv:1206.6074 [hep-ex]].

\bibitem{ATLAS:2012mja}
  ATLAS Collaboration,
  \emph{Search for a Standard Model Higgs in the mass range 200--600 GeV in the channel $H\rightarrow ZZ\rightarrow llqq$ with with the ATLAS detector},
  ATLAS-CONF-2012-017, ATLAS-COM-CONF-2012-030.
  
\bibitem{Micco:2013vma}
  B.~Di Micco [ATLAS Collaboration],
  \emph{Search for the Standard Model Higgs boson in the $H\ (\to WW^{(\ast)})\to \ell\nu\ell\nu, \ell\nu qq$ decay modes with the ATLAS detector},
  PoS ICHEP {\bf 2012} (2013) 044.

\bibitem{Diglio:2014vpa}
  S.~Diglio,
  \emph{Search for a high mass Higgs boson using the ATLAS detector},
  EPJ Web Conf.\  {\bf 71} (2014) 00038.

\bibitem{CMS:2013cda}
  CMS Collaboration,
  \emph{Search for a Standard Model-like Higgs boson decaying into $WW \to l \nu q\bar{q}$ in pp collisions at $\sqrt{s} =$ 8 TeV},
  CMS-PAS-HIG-13-008.

\bibitem{CMS:2015mda}
  CMS Collaboration,
  \emph{Search for a Standard Model-like Higgs boson in the $H \to ZZ \to l^+l^- q \bar{q}$ decay channel at $\sqrt{s}$=8 TeV},
  CMS-PAS-HIG-14-007.

\bibitem{CMS:2015lda}
  CMS Collaboration,
  \emph{Search for a Standard Model-like Higgs boson decaying into $WW \to l \nu q \bar{q}$ in exclusive jet bins in pp collisions at $\sqrt{s}$ = 8 TeV},
  CMS-PAS-HIG-14-008.

\bibitem{Khachatryan:2015cwa}
  V.~Khachatryan {\it et al.}  [CMS Collaboration],
  \emph{Search for a Higgs boson in the mass range from 145 to 1000 GeV decaying to a pair of W or Z bosons},
  arXiv:1504.00936 [hep-ex].

\bibitem{Pelliccioni:2015hva}
  M.~Pelliccioni [CMS Collaboration],
  \emph{CMS high-mass WW and ZZ Higgs search with the complete LHC Run 1 statistics},
  arXiv:1505.03831 [hep-ex].
  

\bibitem{Hackstein:2010wk}
  C.~Hackstein and M.~Spannowsky,
  \emph{Boosting Higgs discovery: the forgotten channel}
  Phys.\ Rev.\ D {\bf 82} (2010) 113012
  [arXiv:1008.2202 [hep-ph]].

  
\bibitem{Dobrescu:2009zf}
  B.~A.~Dobrescu and J.~D.~Lykken,
  \emph{Semileptonic decays of the standard Higgs boson},
  JHEP {\bf 1004} (2010) 083
  [arXiv:0912.3543 [hep-ph]].

\bibitem{Lykken:2011uv}
  J.~D.~Lykken, A.~O.~Martin and J.~C.~Winter,
  \emph{Semileptonic decays of the Higgs boson at the Tevatron}
  JHEP {\bf 1208} (2012) 062
  [arXiv:1111.2881 [hep-ph]].

  

\bibitem{Kao:2012zj}
  C.~Kao and J.~Sayre,
  \emph{Confirming the LHC Higgs discovery with WW},
  Phys.\ Lett.\ B {\bf 722} (2013) 324
  [arXiv:1212.0929 [hep-ph]].

\bibitem{Alwall:2014hca}
  J.~Alwall, R.~Frederix, S.~Frixione, V.~Hirschi, F.~Maltoni, O.~Mattelaer, H.-S.~Shao and T.~Stelzer {\it et al.},
  \emph{The automated computation of tree-level and next-to-leading order differential cross sections, and their matching to parton shower simulations},
  JHEP {\bf 1407} (2014) 079
  [arXiv:1405.0301 [hep-ph]].

\bibitem{Furry:1937zz}
  W.~H.~Furry,
  \emph{A symmetry theorem in the positron theory},
  Phys.\ Rev.\  {\bf 51} (1937) 125.
  
\bibitem{Campbell:2011bn}
  J.~M.~Campbell, R.~K.~Ellis and C.~Williams,
  \emph{Vector boson pair production at the LHC}
  JHEP {\bf 1107} (2011) 018
  [arXiv:1105.0020 [hep-ph]].
  

\bibitem{gg2VV}
\url{http://gg2VV.hepforge.org/}

\bibitem{Hahn:2000kx}          
  T.~Hahn,
  \emph{Generating Feynman diagrams and amplitudes with FeynArts 3},
  Comput.\ Phys.\ Commun.\  {\bf 140} (2001) 418
  [arXiv:hep-ph/0012260].

\bibitem{Hahn:1998yk}             
  T.~Hahn and M.~Perez-Victoria,
  \emph{Automatized one-loop calculations in four and $D$ dimensions},
  Comput.\ Phys.\ Commun.\  {\bf 118} (1999) 153
  [arXiv:hep-ph/9807565].
  
  
\bibitem{Goria:2011wa}
  S.~Goria, G.~Passarino and D.~Rosco,
  \emph{The Higgs boson lineshape},
  Nucl.\ Phys.\ B {\bf 864} (2012) 530
  [arXiv:1112.5517 [hep-ph]].
  
\bibitem{Berends:1994pv}
  F.~A.~Berends, R.~Pittau and R.~Kleiss,
  \emph{All electroweak four fermion processes in electron - positron collisions}
  Nucl.\ Phys.\ B {\bf 424} (1994) 308
  [hep-ph/9404313].

\bibitem{Kauer:2015hia}
  N.~Kauer and C.~O’Brien,
  \emph{Heavy Higgs signal-background interference in $gg\to VV$ in the Standard Model plus real singlet},
  Eur.\ Phys.\ J.\ C {\bf 75} (2015) 374
  [arXiv:1502.04113 [hep-ph]].

\bibitem{Hirschi:2015iia}
  V.~Hirschi and O.~Mattelaer,
  \emph{Automated event generation for loop-induced processes},
  arXiv:1507.00020 [hep-ph].

\bibitem{launchpad}
\url{https://launchpad.net/mg5amcnlo/2.0/2.3.0/+download/MG5_aMC_v2.3.0.tar.gz}

\bibitem{large-scale}
V.~Hirschi, A. ~Kalogeropoulos and  O.~Mattelaer,
\emph{Optimization techniques for large scale Monte Carlo event generation at the LHC},
in preparation.

\bibitem{Hirschi:2011pa}
  V.~Hirschi, R.~Frederix, S.~Frixione, M.~V.~Garzelli, F.~Maltoni and R.~Pittau,
  \emph{Automation of one-loop QCD corrections},
  JHEP {\bf 1105} (2011) 044
  [arXiv:1103.0621 [hep-ph]].
  
\bibitem{Ossola:2006us}
  G.~Ossola, C.~G.~Papadopoulos and R.~Pittau,
  \emph{Reducing full one-loop amplitudes to scalar integrals at the integrand level},
  Nucl.\ Phys.\ B {\bf 763} (2007) 147
  [hep-ph/0609007].
\bibitem{Ossola:2007ax}
  G.~Ossola, C.~G.~Papadopoulos and R.~Pittau,
  \emph{CutTools: a program implementing the OPP reduction method to compute one-loop amplitudes},
  JHEP {\bf 0803} (2008) 042
  [arXiv:0711.3596 [hep-ph]].

\bibitem{Martin:2009iq}
  A.~D.~Martin, W.~J.~Stirling, R.~S.~Thorne and G.~Watt,
  \emph{Parton distributions for the LHC},
  Eur.\ Phys.\ J.\ C {\bf 63} (2009) 189
  [arXiv:0901.0002 [hep-ph]].

\end{thebibliography}
\end{document}